\documentclass[prb,twocolumn]{revtex4-1}
\usepackage{graphicx}

\begin{document}
\title{Gauge Fluctuations and Interlayer Coherence in Bilayer Composite Fermion Metals}

\author{R. Cipri and N.E. Bonesteel}
\affiliation{Department of Physics and NHMFL, Florida State University, Tallahassee, Florida 32310, USA}

\date{\today}

\begin{abstract}
We study the effect of the Chern-Simons gauge fields on the possible transition from two decoupled composite fermion metals to the interlayer coherent composite fermion state proposed by Alicea et al. [Phys. Rev. Lett. {\bf 103}, 256403 (2009)] in a symmetrically doped quantum Hall bilayer with total Landau level filling fraction $\nu_{tot} = 1$.  In this transition, interlayer Coulomb repulsion leads to excitonic condensation of composite fermions which are then free to tunnel coherently between layers.  We find that this coherent tunneling is strongly suppressed by the layer-dependent Aharonov-Bohm phases experienced by composite fermions as they propagate through the fluctuating gauge fields in the system. This suppression is analyzed by treating these gauge fluctuations within the random-phase approximation and calculating their contribution to the energy cost for forming an exciton condensate of composite fermions. This energy cost leads to (1) an increase in the critical interlayer repulsion needed to drive the transition; and (2) a discontinuous jump in the energy gaps to out-of-phase excitations (i.e., excitations involving currents with opposite signs in the two layers) at the transition.  
\end{abstract}


\maketitle

\section{Introduction}

The quantum Hall bilayer with total Landau level filling fraction $\nu_{tot} = 1$ is a particularly rich system for studying quantum Hall physics.\cite{dassarma_book97,eisenstein04}  In this system, two parallel two-dimensional electron gases, separated by a distance $d$, are placed in a perpendicular magnetic field $B$ such that the total electron density of the two layers is that of a filled Landau level for a single layer. For the symmetrically doped case, each layer then has Landau level filling fraction $\nu=1/2$. If interlayer electron tunneling can be ignored, the only coupling between layers is through the Coulomb repulsion.  The scale of this coupling, relative to the scale of interactions within each layer, is then set by the dimensionless ratio $d/l_0$, where $l_0 = (\hbar c/(eB))^{1/2}$ is the magnetic length. 

In the limit of small $d/l_0$ (strong interlayer coupling) this system enters a remarkable bilayer quantum Hall state in which electrons develop spontaneous interlayer phase coherence.\cite{moon95} This state can be viewed as an exciton condensate formed by electron-hole pairs in the two layers.\cite{dassarma_book97,eisenstein04}  In the opposite limit of large $d/l_0$ (weak interlayer coupling) the correlations within each layer presumably give rise to two separate $\nu=1/2$ composite fermion metals, compressible states in which physical electrons are represented by new particles, composite fermions, attached to two fictitious (Chern-Simons) flux quanta.\cite{jain89,jainbook,halperin93}  These composite fermions then move in zero effective magnetic field, forming two Fermi surfaces, one in each layer.\cite{halperin93} Despite a great deal of experimental\cite{kellogg03,spielman05,kumada05,luin05,giudici08,karmakar09,finck10,giudici10} and theoretical\cite{cote92,bonesteel93,bonesteel96,kim01,schliemann01,stern02,simon03,shibata06,moeller08,moeller09,milovanovic09,zou10} work devoted to studying the crossover between these two limiting cases, the nature of this crossover is still poorly understood.
 
Alicea et al.\cite{alicea09} have made the interesting proposal that short-range interlayer repulsion in the $\nu_{tot}=1$ bilayer could lead to a state for intermediate $d/l_0$ in which composite fermions, rather than physical electrons, undergo excitonic condensation and thus develop spontaneous interlayer phase coherence.  The starting point for understanding this interlayer coherent composite fermion state is the large $d/l_0$ limit of two decoupled composite fermion metals. As $d/l_0$ is decreased, the interlayer Coulomb repulsion grows and, when strong enough, can lead to excitonic condensation of composite fermions. If this occurs, the composite fermions become liberated from their layers and are able to tunnel coherently between them, even though physical electrons do not.  This tunneling leads to the formation of well-defined bonding and antibonding composite fermion bands that are split in energy, with one composite fermion Fermi surface growing and the other shrinking.  As shown in Ref.~\onlinecite{alicea09}, the resulting state is compressible in the in-phase sector and incompressible in the out-of-phase sector, where excitations in the in-phase (out-of-phase) sector involve currents with the same (opposite) sign in the two layers.  Incompressibility in the out-of-phase sector then implies a quantized Hall effect in the counterflow channel.  Despite plausible arguments for the favorability of this state over a range of $d/l_0$,\cite{alicea09} there is currently no experimental evidence for it forming in $\nu_{tot}=1$ quantum Hall bilayers.  One purpose of the present work is to provide a possible explanation for this.

In this paper we study the effect of the gauge fields associated with the Chern-Simons flux attached to composite fermions on the transition from two decoupled composite fermion metals to an interlayer coherent composite fermion state.  These gauge fields lead to strongly fluctuating layer-dependent Aharonov-Bohm phases experienced by composite fermions as they propagate through the system, and so it is natural to suspect they will strongly suppress any interlayer coherence these composite fermions may have. Here we find that this is, in fact, the case.  Our analysis is based on treating the fluctuating gauge fields within the random-phase approximation (RPA) and calculating their contribution to the energy cost for forming an interlayer coherent composite fermion state.  This energy cost results in a significant increase in the interlayer repulsion strength required to drive the transition to this state.  In addition, the energy gaps to out-of-phase excitations, which are found to open continuously at the transition when gauge fluctuations are ignored, jump discontinuously at the transition when gauge fluctuations are included.

This paper is organized as follows.  In Sec.~\ref{secII} we review the bilayer model studied in Ref.~\onlinecite{alicea09} and the mean-field theory of the transition from two decoupled composite fermion metals to the interlayer coherent composite fermion state. In Sec.~\ref{secIII} we argue that gauge fluctuations should play an important role in determining the nature of this transition and carry out an RPA analysis of these fluctuations.  This analysis allows us to calculate how the collective modes of the system are affected by the formation of the interlayer coherent composite fermion state.  We then compute the RPA contribution to the correlation energy in this state due to gauge fluctuations and analyze the effect this contribution has on the transition.  Finally, our conclusions are summarized in Sec.~\ref{secIV}.

\section{Spontaneous Interlayer Phase Coherence of Composite Fermions}
\label{secII}

We consider the idealized case of a disorder free, fully spin-polarized,\cite{spin_note} symmetrically doped bilayer with zero interlayer tunneling and total filling fraction $\nu_{tot}=1/p$ where $p$ is an integer.  Each layer then has even denominator filling fraction $\nu=1/(2p)$. When these layers are well-separated we assume that each can be described as a composite fermion metal. In this description physical electrons are represented as composite fermions with $2p$ Chern-Simons flux quanta attached to them,\cite{jain89,jainbook,halperin93} where the flux attached to particles in a given layer is seen only by composite fermions in that same layer.\cite{bonesteel93}  At the mean-field level, the fictitious magnetic field associated with this flux exactly cancels the applied magnetic field and the composite fermions in each layer move in zero effective magnetic field.\cite{halperin93}

The specific model studied here was introduced in Ref.~\onlinecite{alicea09}. In this model it is assumed that the primary role of the Coulomb interaction within each layer is to induce the formation of the two composite fermion metals.  The only interaction included explicitly is then an interlayer delta-function repulsion, $u\delta({\bf r}_1 - {\bf r}_2)$, meant to describe the short-range part of the Coulomb interaction between layers. The Euclidean-time action for this model at temperature $T$ is $S = \int_0^\beta d \tau \int d^2 r {\cal L}({\bf r},\tau)$ where $\beta = (k_B T)^{-1}$ and the Lagrangian density is ${\cal L} = {\cal L}_0 + {\cal L}_{int} + {\cal L}_{CS}$ with
\begin{eqnarray}
{\cal L}_0 = \sum_{\alpha=\uparrow,\downarrow} \overline{\psi}_\alpha \Bigl(\partial_\tau - i a_0^\alpha   - \frac{1}{2m^*}(\nabla
- i {\bf a}^\alpha)^2 \Bigr) \psi_\alpha,
\label{L0}
\end{eqnarray}
\begin{eqnarray}
{\cal L}_{int} = u \overline{\psi}_\uparrow\overline{\psi}_\downarrow\psi_\downarrow \psi_\uparrow,
\label{Lint}
\end{eqnarray}
and
\begin{eqnarray}
{\cal L}_{CS} = -\frac{i}{2\pi\lambda} \sum_{\alpha = \uparrow,\downarrow} a_0^\alpha {\hat {\bf z}} \cdot \left(\nabla \times \left({\bf a}^\alpha + e{\bf A}_{ext}\right)\right).
\label{LCS}
\end{eqnarray}
Here $\psi_\alpha$ is the composite fermion field in layer $\alpha$ where $\alpha = \uparrow,\downarrow$ is a pseudospin label for the layers, ${\bf A}_{ext}$ is the vector potential for the external applied magnetic field ${\bf B} = \nabla \times {\bf A}_{ext} = {\bf \hat z} 2\pi\lambda n/e$ where $n$ is the electron density in each layer, $\lambda = 2p$ is the number of flux quanta attached to each composite fermion ($\lambda = 2$ for the case $\nu_{tot}=1$), $m^*$ is the effective mass of the composite fermions, and $(a_0^\alpha, {\bf a}^\alpha)$ is the Chern-Simons gauge field seen by composite fermions in layer $\alpha$ (here, and in what follows, we take $\hbar = c = 1$).  ${\cal L}_{CS}$ is a Chern-Simons term in the Coulomb gauge for which $\nabla \cdot {\bf a}^\alpha = 0$.   The only gauge degrees of freedom in each layer are then the time component, $a^\alpha_0$, and (after Fourier transforming to momentum space) the transverse component, $a^\alpha_1({\bf q},\tau) = {\bf {\hat z}} \cdot ({\bf {\hat q}} \times {\bf a}^\alpha ({\bf q}, \tau))$, of the Chern-Simons gauge fields.  The partition function is then ${\cal Z} = \int \prod_{\alpha=\uparrow,\downarrow} D\psi_\alpha D a^\alpha_0 D a^\alpha_1 e^{-S}$. 

Integrating out the time components of the Chern-Simons gauge fields enforces the constraint $\nabla \times {\bf a}^\alpha = \hat{\bf z} 2\pi\lambda  \delta \rho_\alpha$ where $\delta\rho_\alpha = \bar\psi_\alpha \psi_\alpha - n$ is the fluctuation of the density in layer $\alpha$ about its mean value.  Gauge field fluctuations in each layer are thus tied to density fluctuations in that layer.  As a first approximation, if we ignore these fluctuations (and so set $a_0^\alpha=0$ and ${\bf a}^\alpha=0$), then, at the mean-field level, the instability to the interlayer coherent composite fermion state discussed by Alicea et al.\cite{alicea09} is a simple Stoner instability.  In pseudospin language the instability is to a pseudospin ferromagnet in which the layer pseudospins are polarized along a certain direction in the $xy$ plane.  A similar instability to the formation of spontaneous interlayer coherence for electrons in bilayers in zero magnetic field was studied in Ref.~\onlinecite{zheng97}.

\begin{figure}[t]
     \includegraphics[width=2.7in]{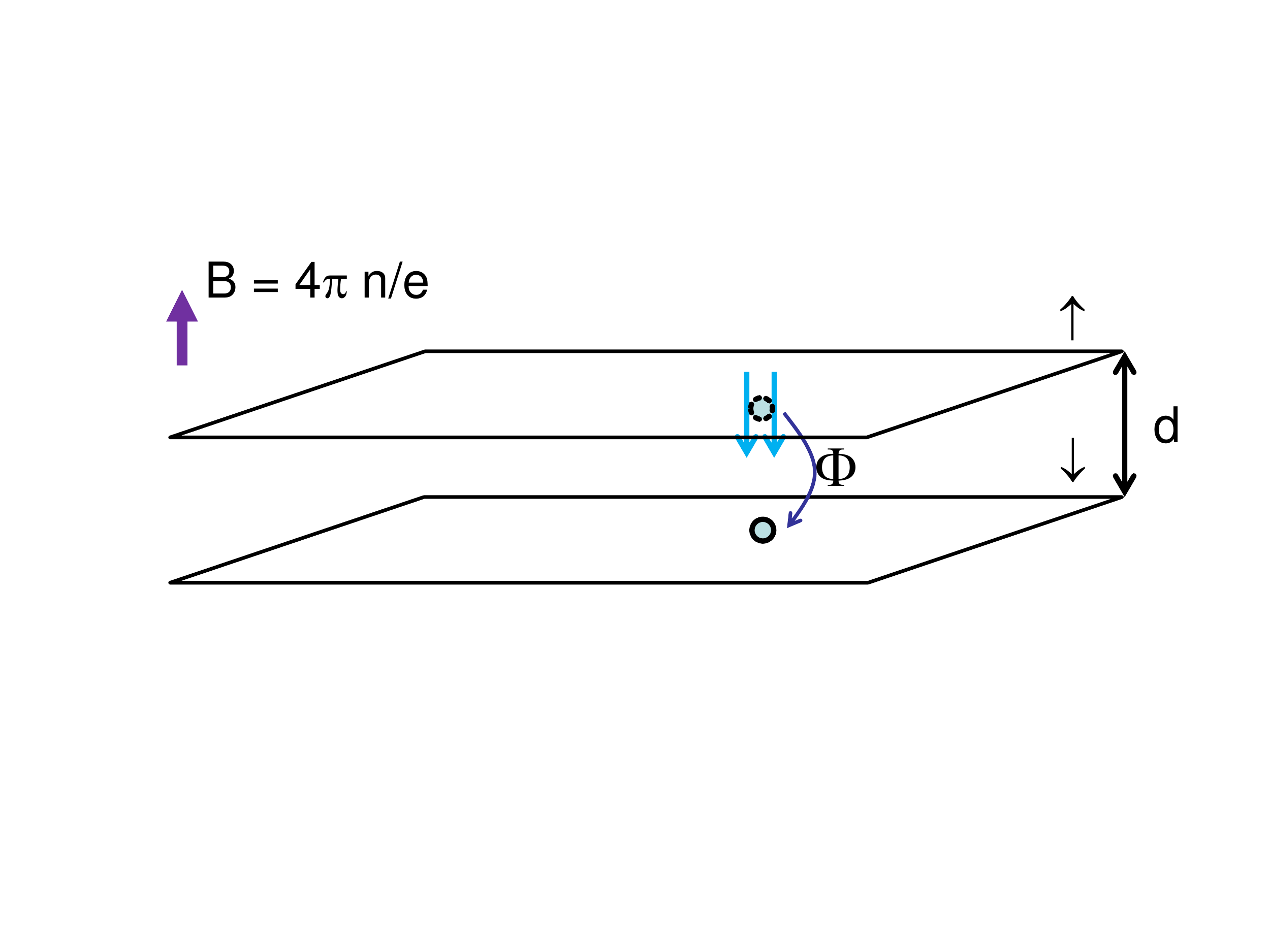}			
\caption{(Color online). Symmetrically doped $\nu_{tot}=1$ quantum Hall bilayer (the case $\lambda =2$ in the text).  Layers are labeled by pseudospin indices $\uparrow$ and $\downarrow$. $d$ is the layer spacing and $B = 4\pi n/e$  is the external magnetic field, where $n$ is the electron density in each layer ($\hbar = c = 1$). The filling factor in each layer is $\nu = 1/2$ and electrons are represented as composite fermions bound to two flux quanta, as shown in the top layer.  The interlayer coherent composite fermion state proposed in Ref.~\onlinecite{alicea09} is characterized by a nonzero interlayer tunneling amplitude $\Phi$ for composite fermions even though there is no interlayer tunneling for electrons.} \label{bilayer}
 \end{figure}

This instability can be studied by first carrying out a Hubbard-Stratonovich transformation.  This is accomplished by multiplying the partition function ${\cal Z}$ by the constant factor $\int D\Phi~ e^{-S_{HS}}$, where $S_{HS} = \int_0^\beta d\tau \int d^2 {\bf r} {\cal L}_{HS}$, with
\begin{eqnarray}
{\cal L}_{HS} = \overline{c}({\bf r},\tau) c({\bf r},\tau),
\label{hs1}
\end{eqnarray}
and
\begin{eqnarray}
c({\bf r},\tau) = \frac{1}{\sqrt{u}} \Phi({\bf r},\tau) - \sqrt{u}~ \overline{\psi}_\downarrow ({\bf r},\tau) \psi_\uparrow ({\bf r},\tau).
\label{hs2}
\end{eqnarray}
${\cal L}_{HS}$ can then be added to the Lagrangian density for the interlayer interaction to give,
\begin{eqnarray}
{\cal L}_{int} + {\cal L}_{HS} = \frac{1}{u} |\Phi|^2 - \Phi~\overline{\psi}_\uparrow \psi_\downarrow
- \Phi^*~\overline{\psi}_\downarrow \psi_\uparrow.
\label{Linths}
\end{eqnarray}

At the mean-field level we take the Hubbard-Stratonovich field $\Phi$ to be uniform in space and constant in time.  $\Phi$ is then the order parameter for the interlayer coherent composite fermion state and acts as a fixed effective interlayer tunneling amplitude for composite fermions (see Fig.~\ref{bilayer}).  Without loss of generality we take $\Phi$ to be real and positive in what follows. The system is then diagonalized by the following change of variables to fields which describe composite fermions in symmetric ($S$) and antisymmetric ($A$) interlayer states,
\begin{eqnarray}
\psi_S &=& \left(\psi_\uparrow + \psi_\downarrow\right)/\sqrt{2},\label{psiS}\\
\psi_A &=& \left(\psi_\uparrow - \psi_\downarrow\right)/\sqrt{2}.\label{psiA}
\end{eqnarray}
After this transformation the mean-field Lagrangian density becomes
\begin{eqnarray}
{\cal L}_{MF} &=&  \frac{\Phi^2}{u^2} +  \overline{\psi}_S\left(\partial_\tau -\Phi - \frac{1}{2m^*}\nabla ^2 \right) \psi_S\nonumber\\
&&~~~~~ + \overline{\psi}_A\left(\partial_\tau + \Phi - \frac{1}{2m^*}\nabla ^2 \right) \psi_A.
\label{LMF}
\end{eqnarray}
Fourier transforming from real space to momentum space then yields the dispersions of the symmetric and antisymmetric bands, which are simply those of noninteracting particles shifted by $\pm \Phi$ 
\begin{eqnarray}
{\cal E}_{\bf k}^{S} &=& {\cal E}_{\bf k} - \Phi,\label{dispS}\\
{\cal E}_{\bf k}^{A} &=& {\cal E}_{\bf k} + \Phi,\label{dispA}
\label{eq:}
\end{eqnarray}
where ${\cal E}_{\bf k} = k^2/(2m^*)$.  

\begin{figure}[t]
  \begin{center}
    \includegraphics[width=\columnwidth]{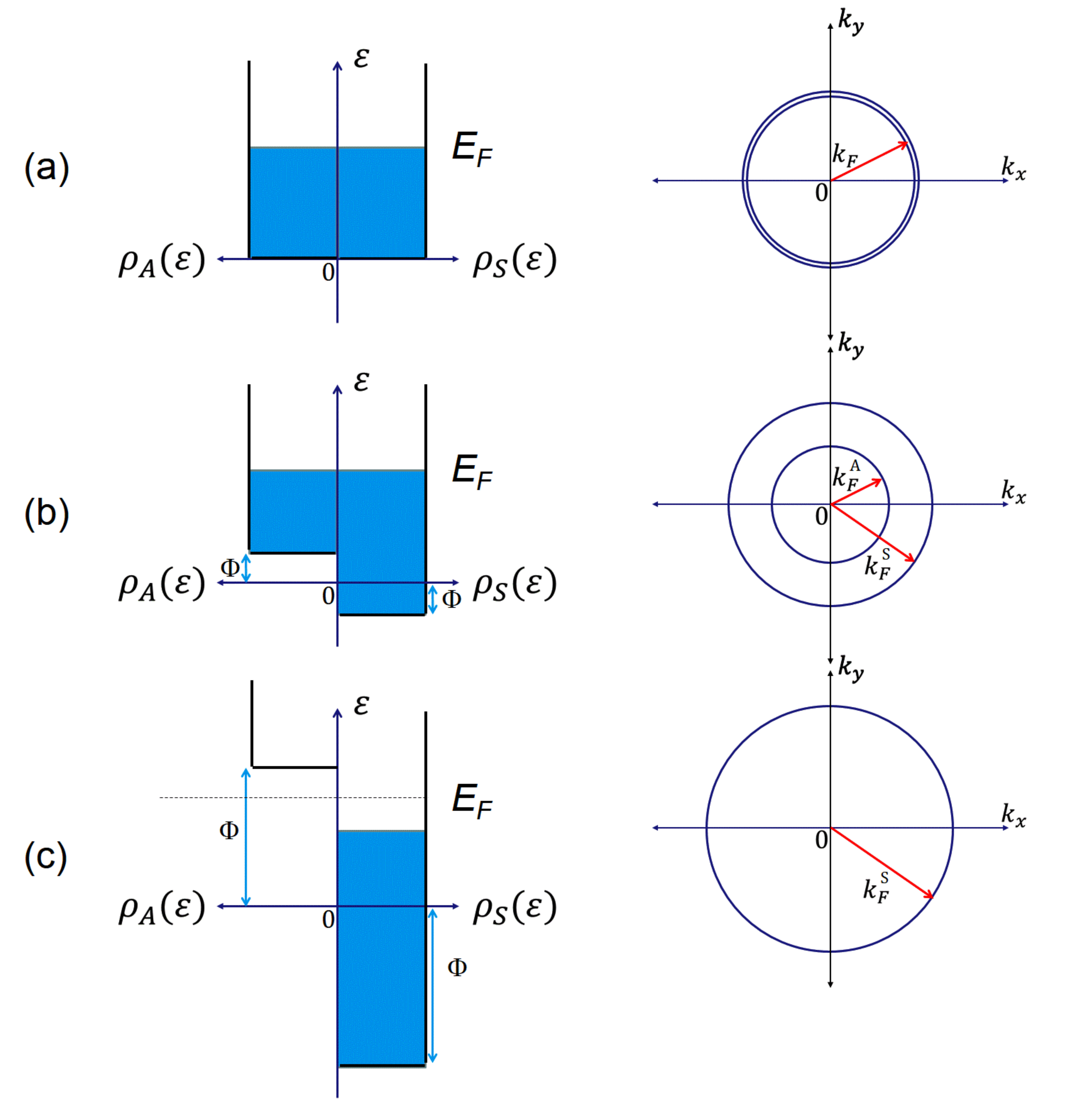}
		\caption{(Color online). Density of states and Fermi surfaces for the symmetric and antisymmetric composite fermion bands when (a) $\Phi = 0$, (b) $0 < \Phi < E_F$, and (c) $\Phi > E_F$.   Here $\rho_S$ and $\rho_A$ are, respectively, the densities of states for the symmetric and antisymmetric band and $E_F$ is the Fermi energy for $\Phi = 0$.  The Fermi energy remains fixed and equal to $E_F$ for $0 < \Phi < E_F$ and is equal to $2E_F - \Phi$ for $\Phi > E_F$.  Expressions for $k_F^S$ and $k_F^A$ are given in the text.} \label{dos}
   \end{center}
\end{figure}

Because of this splitting, the Fermi surfaces for symmetric and antisymmetric composite fermions have different Fermi wavevectors. If we define $E_F = k_F^2/(2m^*)$ to be the Fermi energy when $\Phi = 0$ then the two Fermi wavevectors for $\Phi < E_F$ are,
\begin{eqnarray}
k_F^S &=& k_F \left(1+\frac{\Phi}{E_F}\right)^{1/2},\label{kfS}\\
k_F^A &=& k_F \left(1-\frac{\Phi}{E_F}\right)^{1/2}.\label{kfA}
\end{eqnarray}
As $\Phi$ increases from 0 the Fermi energy is initially fixed at $E_F$. There are then two Fermi surfaces, and $k_F^S$ increases while $k_F^A$ decreases until, when $\Phi = E_F$ and for $\Phi > E_F$, the Fermi energy is given by $2 E_F - \Phi$, $k_F^A = 0$ and there is only a single Fermi surface with Fermi wavevector $k_F^S = \sqrt{2} k_F$. The density of states for the symmetric and antisymmetric bands and the corresponding Fermi surfaces when $\Phi = 0$, $0 < \Phi < E_F$, and $\Phi > E_F$ are shown in Fig.~\ref{dos}.

Upon integrating out the composite fermion fields while keeping $\Phi$ fixed and taking the $T \rightarrow 0$ limit of the free energy $F = -\beta^{-1} \ln {\cal Z}$ one obtains the following expression for the ground state energy density as a function of $\Phi$ measured with respect to the energy density of the system when $\Phi = 0$,
\begin{eqnarray}
\frac{E_S(\Phi)}{\nu_0 E_F^2} = \left\{
\begin{array}{cl}
\left(\frac{1}{g}-1\right)\frac{\Phi^2}{E_F^2},& \Phi < E_F,\\
\left(\frac{1}{g}-1\right)\frac{\Phi^2}{E_F^2} + 
\left(\frac{\Phi}{E_F}-1\right)^2,& \Phi> E_F.
\end{array}\right.
\label{estoner}
\end{eqnarray}
Here $\nu_0 = m^*/(2\pi)$ is the density of states per band and 
$g = u \frac{m^*}{2\pi} = u \nu_0$ is a dimensionless coupling constant.  The two cases in (\ref{estoner}) correspond to either having two Fermi surfaces ($\Phi < E_F$) or a single Fermi surface ($\Phi > E_F$). The fact that the energy density is a purely quadratic function of $\Phi$ for $\Phi < E_F$ is due to the flat density of states for free particles in two dimensions.  

\begin{figure}[t]
  \begin{center}
   \includegraphics[width=\columnwidth]{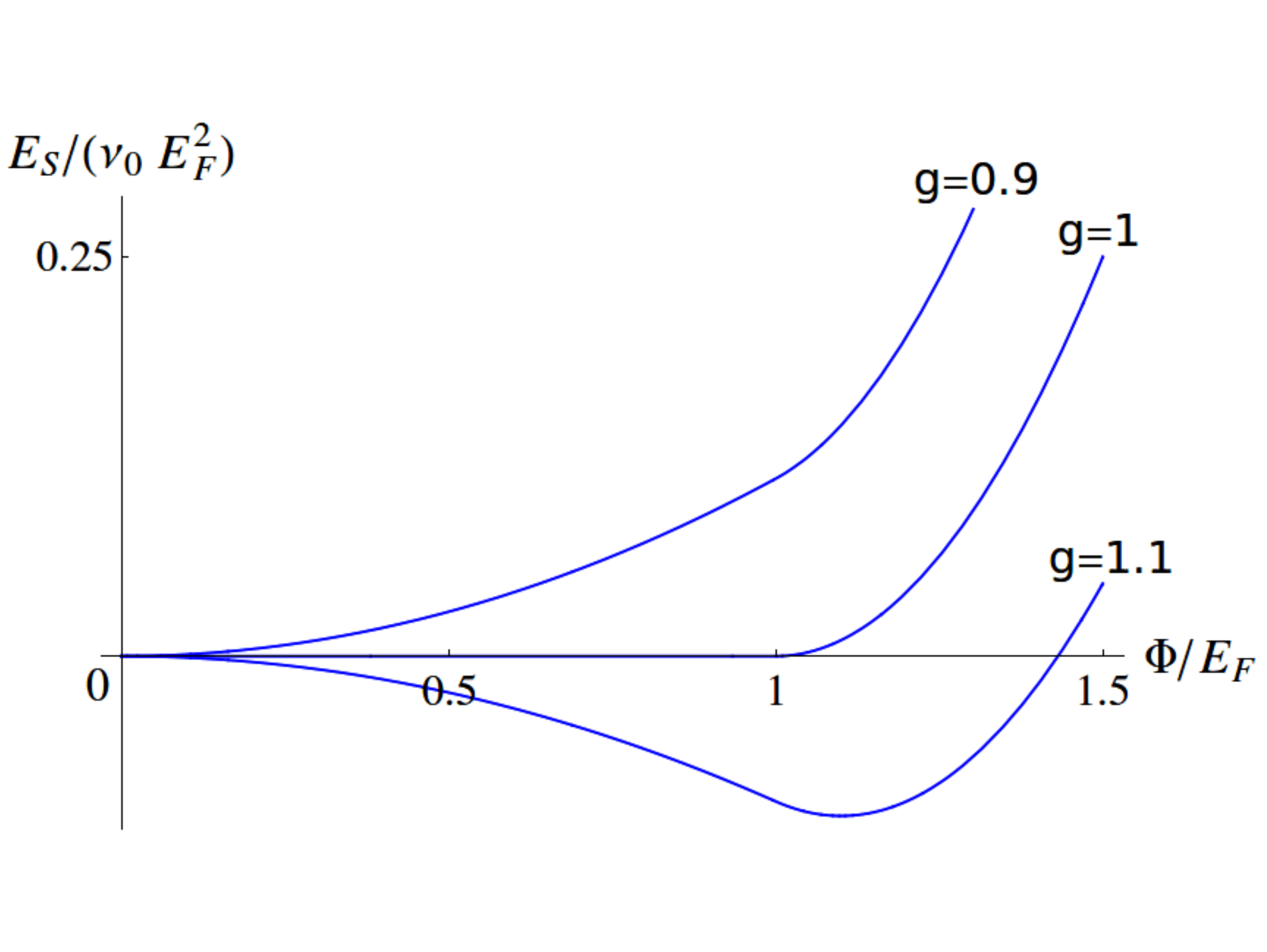}
  \caption{(Color online). Energy density as a function of the order parameter $\Phi$ for different values of the coupling constant $g$ showing the mean field Stoner instability to the interlayer coherent composite fermion state.  Here $g =1$ is the critical value of the coupling constant and the plots are for $g=0.9, 1$, and $1.1$. } \label{stoner}
   \end{center}
\end{figure}

At this level of approximation the Stoner instability occurs when $g = 1$, as shown in Fig.~\ref{stoner}.  For $g<1$ the energy density, $E_S(\Phi)$, is minimized for $\Phi = 0$.  At the critical point, $g=1$ and $E_S(\Phi)$ is independent of $\Phi$ for $\Phi < E_F$. Then, for $g=1+\epsilon$, the order parameter minimizing $E_S(\Phi)$ jumps from $\Phi = 0$ to $\Phi = E_F$, signaling the formation of an exciton condensate of composite fermions with $\langle \bar\psi_\uparrow \psi_\downarrow \rangle \ne 0$, and establishing the interlayer coherent composite fermion state.  We see that the transition is directly to a {\it fully} polarized pseudospin ferromagnet, i.e., a state in which {\it all} composite fermions are in the symmetric band.  This is a consequence of the purely quadratic behavior of $E_{S}(\Phi)$ for $\Phi < E_F$ due to the flat density of states described above.   Note that for $g=1+\epsilon$ the energy gap for interband particle-hole excitations is zero; when $\Phi = E_F$ there are zero energy excitations with wavevector $q = \sqrt{2} k_F$ in which a composite fermion is promoted from the Fermi surface of the symmetric band at $k_F^S = \sqrt{2} k_F$ to the bottom of the (empty) antisymmetric band at $k=0$.  For $\Phi > E_F$ an energy gap, $\Delta_{q=\sqrt{2} k_F}$, opens for these $q=\sqrt{2} k_F$ interband excitations where,
\begin{eqnarray}
\Delta_{q=\sqrt{2} k_F} = 2(\Phi -E_F).\label{deltaqs2}
\end{eqnarray}
For $g>1$, the Stoner energy $E_{S}(\Phi)$ is minimized when $\Phi = g E_F$ and so this gap opens continuously at the transition as $\Delta_{q=\sqrt{2} k_F} = 2(g-1)E_F$.

The simplified model considered here is best viewed as an effective low-energy theory for a bilayer composite fermion metal. Here and in what follows we take this model at face value, particularly because, as we will see in the next Section, the RPA analysis of gauge fluctuations can be carried out essentially analytically.  Following Alicea et al.\cite{alicea09} we can use the renormalized effective mass $m^* \simeq 6/(e^2 l_0)$ from Ref.~\onlinecite{murthy03} for $\lambda = 2$.  This effective mass is set by the intralayer Coulomb interaction energy (the only energy scale in the lowest Landau level) and is independent of the bare band mass of the electrons.  If, also following Alicea et al.,\cite{alicea09} we take $u \simeq (e^2/d)(\pi l_0^2)$ to roughly model the short-range part of the interlayer Coulomb interaction, then the dimensionless coupling constant is $g \simeq 3 l_0/d$ and the critical layer spacing for the Stoner instability is $\left(d/l_0\right)_c \simeq 3$. 

\section{Effect of Gauge Fluctuations}
\label{secIII}

The Stoner instability analysis described in the previous section does not take into account the effect of fluctuations in the Chern-Simons gauge fields attached to the composite fermions.  There are good reasons for thinking these fluctuations will be important.  Fluctuations in these gauge fields lead to fluctuations in the Aharonov-Bohm phases experienced by composite fermions.  Because these fluctuations are different in the two layers, any interlayer phase coherence these composite fermions may have will quickly be lost as a they propagate through this wildly fluctuating ``gauge sea.'' 

\begin{figure}[t]
  \begin{center}    \includegraphics[width=\columnwidth]{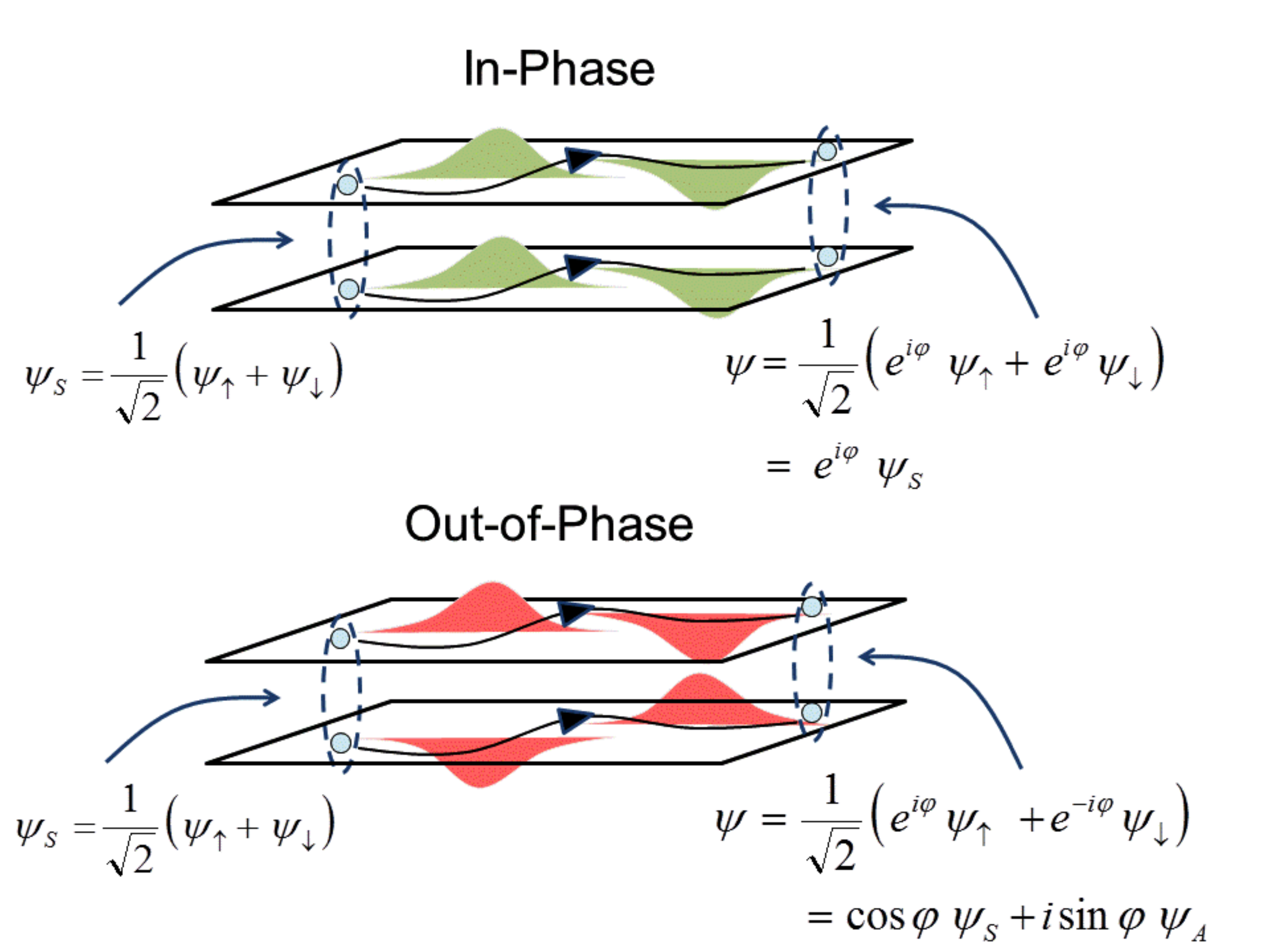}
\caption{(Color online). Effect of in-phase and out-of-phase gauge fluctuations on a composite fermion as it propagates through the bilayer starting in the symmetric band.  In-phase gauge fluctuations are tied to in-phase density fluctuations, represented schematically in green.  These fluctuations give the propagating composite fermion a layer-independent Aharonov-Bohm phase $\phi$.  This phase does not affect interlayer coherence and leads only to intraband scattering within the $S$ and $A$ bands. Out-of-phase gauge fluctuations are likewise tied to out-of-phase density fluctuations, shown in red. These fluctuations give the propagating composite fermion opposite Aharonov-Bohm phases $\pm \phi$ in the two layers.  These fluctuations strongly inhibit interlayer phase coherence and lead to interband scattering between the $S$ and $A$ bands.} \label{gauge}
   \end{center}
\end{figure}

This effect can be seen clearly by introducing in-phase ($a^+$) and out-of-phase ($a^-$) gauge fields,\cite{bonesteel93}
\begin{eqnarray}
a^+_\mu &=& (a^\uparrow_\mu + a^\downarrow_\mu)/\sqrt{2}, \label{ap}\\
a^-_\mu &=& (a^\uparrow_\mu - a^\downarrow_\mu)/\sqrt{2}. \label{am}
\end{eqnarray}
Figure \ref{gauge} shows the effect fluctuations in $a^+$ and $a^-$ have on a composite fermion as it propagates through the bilayer.  Assume the composite fermion starts in either a symmetric state $\psi_S$ or antisymmetric state $\psi_A$ (Fig.~\ref{gauge} shows the $\psi_S$ case).  As this composite fermion moves, in-phase gauge fluctuations result in the same Aharonov-Bohm phase regardless of which layer the composite fermion is in.  Thus these fluctuations do not suppress interlayer coherence; a composite fermion that starts in either the symmetric or antisymmetric band will stay in that band as it scatters off of fluctuations of $a^+$.  By contrast, the out-of-phase gauge fluctuations give opposite Aharonov-Bohm phases to composite fermions in layer $\uparrow$ and layer $\downarrow$. Fluctuations in $a^-$ therefore strongly suppress interlayer coherence and lead to interband scattering between the symmetric and antisymmetric bands.  It is interesting to note that while fluctuations in $a^-$ suppress interlayer coherence of composite fermions in the particle-hole channel, these same fluctuations are known to enhance interlayer BCS pairing of composite fermions in the particle-particle channel.\cite{bonesteel93,bonesteel96} 

\begin{figure}
  \begin{center}
\includegraphics[width=\columnwidth]{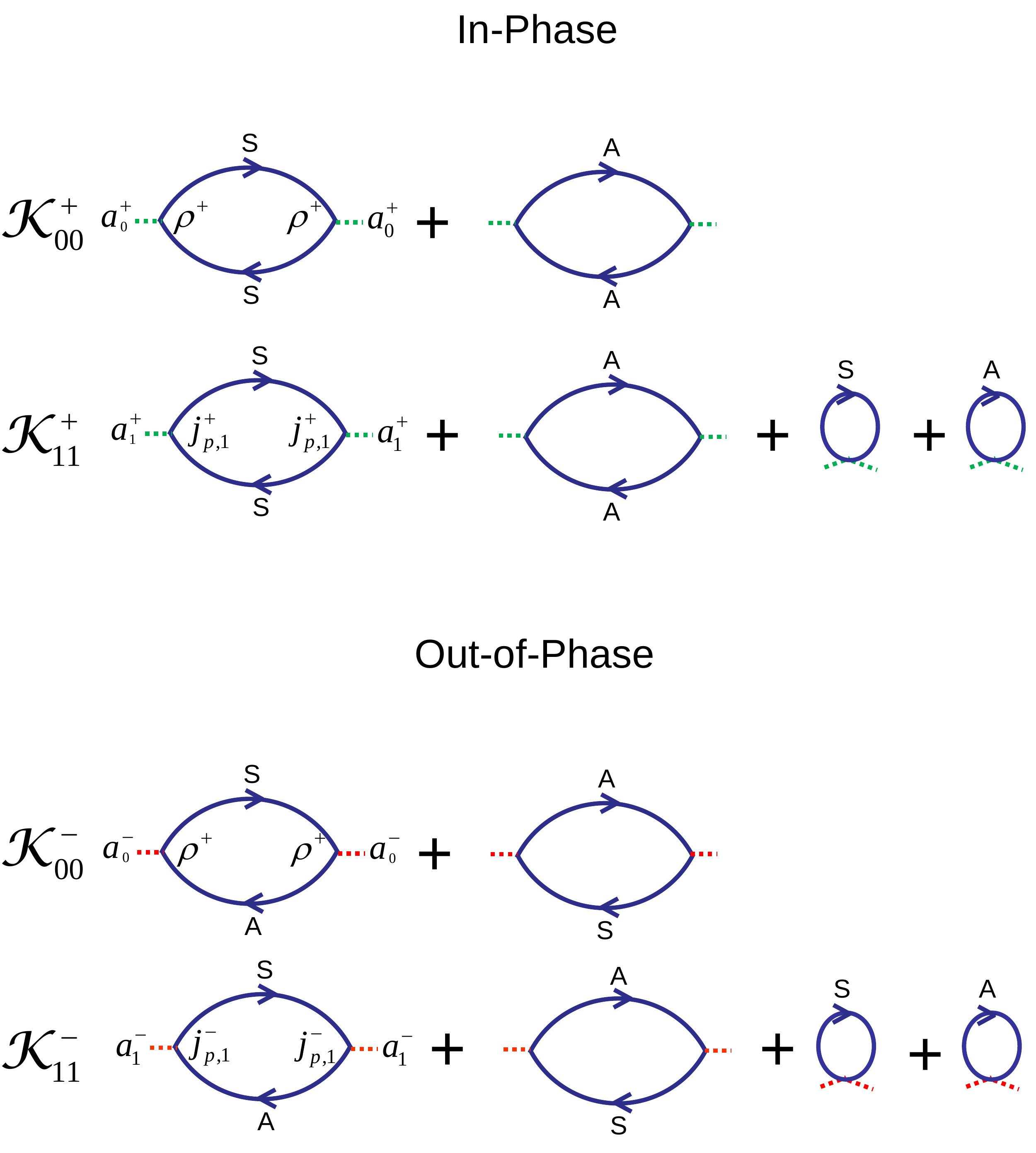}
  \caption{(Color online). Feynman diagrams for ${\cal K}_{00}^\pm({\bf q},\i\omega;\Phi)$ and ${\cal K}_{11}^\pm({\bf q},i\omega;\Phi)$.  $S$ and $A$ label composite fermion propagators (blue) in the symmetric $(+)$ and antisymmetric $(-)$ bands, respectively.  In-phase gauge fields (green) lead to intraband scattering $(S\leftrightarrow S)$, $(A\leftrightarrow A)$, while out-of-phase gauge fields (red) lead to interband scattering $(S \leftrightarrow A)$ (see Fig.~\ref{gauge}).  The seagull diagrams only contribute to ${\cal K}_{11}^\pm$ where they give the diamagnetic contribution $-n/m^*$.} \label{response}
   \end{center}
\end{figure}

The suppression of interlayer coherence by $a^-$ fluctuations is similar to the suppression of BCS pairing of composite fermions in a single-layer $\nu=1/2$ system studied in Ref.~\onlinecite{bonesteel99}.  The main result of this earlier work was the observation that, while in an ordinary BCS transition {\it any} attractive interaction strength, however small, is sufficient for a pairing instability to occur, when the effect of the gauge fluctuations are included a finite interaction strength is required to induce a transition.  This resistance to pairing can be understood as a consequence of singular pair breaking due to the strongly fluctuating effective magnetic field seen by the composite fermions.\cite{bonesteel99} The role of similar gauge fluctuations in preventing the Kohn-Luttinger pairing instability of the Fermi surface in three dimensions has been studied in Ref.~\onlinecite{schafer06} (in the context of high-density quantum chromodynamics) and, more recently, in Ref.~\onlinecite{chung13}.  The Stoner instability studied here differs from BCS paring in that a finite coupling strength is required for the transition to occur even in the absence of gauge fluctuations.  However, as we will see below, the inclusion of gauge fluctuations leads to similar qualitative changes in the nature of the transition.  Thus the model studied here provides another example of the nontrivial effect gauge fluctuations can have on phase transitions in dense Fermi systems.

To analyze the effect of gauge fluctuations on the interlayer coherent composite fermion state within the RPA we begin with the full action defined in Sec.~\ref{secII}, integrate out the fermions for fixed constant $\Phi$, and expand the resulting effective action to second order in $a^+$ and $a^-$.  This expanded action decouples into in-phase and out-of-phase sectors,\cite{bonesteel99} and has the form $S_{RPA} = S_{RPA}^+ + S_{RPA}^-$ where,
\begin{eqnarray}
S^\pm_{RPA} &=& \frac{1}{2} \sum_{\omega_n} \sum_{\bf q} \nonumber \\
&&\sum_{{{\mu=0,1\atop{\nu=0,1}}}} {a^\pm_\mu}^*({\bf q},i\omega_n) {{{\cal D}^\pm}}^{-1}_{\mu\nu} ({\bf q},i\omega_n;\Phi) a^\pm_\nu({\bf q},i\omega_n).\nonumber \\ 
\label{RPA_action}
\end{eqnarray}
Here, as in Sec.~\ref{secII}, $a^\pm_0({\bf q},i\omega_n)$ and $a^\pm_1({\bf q},i\omega_n) = {\bf{\hat z}} \cdot ({\bf \hat q} \times {\bf a}^\pm({\bf q},i\omega_n))$ are, respectively, the time and transverse components of the gauge fields,
\begin{eqnarray}
{{\cal D}^\pm}^{-1}({\bf q},i\omega;\Phi) = \left(\begin{array}{cc} {\cal K}_{00}^{\pm}({\bf q},i \omega;\Phi) & i q/(2\pi\lambda) \\
-i q/(2\pi\lambda) & {\cal K}_{11}^{\pm}({\bf q},i \omega;\Phi)\end{array} \right),\nonumber\\
\label{Dinverse}
\end{eqnarray}
is the inverse of the $2\times 2$ matrix formed by the gauge field propagators evaluated on the imaginary frequency axis, ${\cal D}_{\mu\nu}^{\pm}({\bf q},i\omega_n) = \langle {a^{\pm}_\mu}^*({\bf q},i\omega_n) a^{\pm}_\nu({\bf q},i\omega_n)\rangle$ where $\langle \cdots \rangle = {\cal Z}^{-1}\int D a_0^\alpha a_1^\alpha \cdots e^{-S_{RPA}}$, and $\omega_n = 2n\pi/\beta$ is the $n$th bosonic Matsubara frequency.  

The functions ${\cal K}_{00}^{\pm}({\bf q},i\omega;\Phi)$ and ${\cal K}_{11}^{\pm}({\bf q},i\omega;\Phi)$ appearing in the expression for ${{\cal D}^\pm}^{-1}$ are obtained by evaluating the Feynman diagrams shown in Fig.~\ref{response}.  The vertices in the bubble diagrams for ${\cal K}_{00}^\pm$ and ${\cal K}_{11}^\pm$ correspond, respectively, to the density $\rho^\pm = \rho_\uparrow \pm \rho_\downarrow$ and transverse paramagnetic current (in momentum space) $j_{p,1}^\pm({\bf q},\tau) = {\hat {\bf z}}\cdot({\hat {\bf q}} \times ({\bf j}_{p,\uparrow}({\bf q},\tau) \pm {\bf j}_{p,\downarrow}({\bf q},\tau)))$ where ${\bf j}_{p,\alpha}({\bf q},\tau) = \frac{1}{2m^*}\sum_{{\bf k}} (2{\bf k} + {\bf q}) {\bar\psi}_\alpha({\bf q} + {\bf k},\tau) \psi_\alpha({\bf k},\tau)$, in the in-phase ($+$) and out-of-phase  ($-$) sectors.  We then find
\begin{eqnarray}
{\cal K}^\pm_{\mu\nu}({\bf q},i\omega;\Phi) = \Pi^\pm_{\mu\nu}({\bf q},i\omega;\Phi) - \delta_{\mu,1}\delta_{\nu,1} \frac{n}{m*},\label{kpm}
\end{eqnarray}
where $n= k_F^2/(4\pi)$ is the electron density per layer,
\begin{eqnarray}
\Pi^{+}_{\mu\nu}({\bf q},i\omega;\Phi) = \frac{1}{2} \left(\Pi^{SS}_{\mu\nu}({\bf q},i\omega;\Phi) +
\Pi^{AA}_{\mu\nu}({\bf q},i\omega;\Phi)\right),\nonumber\\
\label{pi+}
\end{eqnarray}
and
\begin{eqnarray}
\Pi^{-}_{\mu\nu}({\bf q},i\omega;\Phi)  = \frac{1}{2} \left(\Pi^{SA}_{\mu\nu}({\bf q},i\omega;\Phi) + \Pi^{AS}_{\mu\nu}({\bf q},i\omega;\Phi)\right).\nonumber\\
\label{pi-}
\end{eqnarray}
Here
\begin{eqnarray}
\Pi^{\alpha\beta}_{00}({\bf q},i\omega;\Phi) = \int \frac{d^2k}{(2\pi)^2} 
\frac{f({\cal E}^\alpha_{{\bf k} + {\bf q}}) - f({\cal E}^\beta_{\bf k})}{i\omega - {\cal E}^\alpha_{{\bf k} + {\bf q}} + {\cal E}^\beta_{\bf k}},
\label{pi00}
\end{eqnarray}
\begin{eqnarray}
\Pi^{\alpha\beta}_{11}({\bf q},i\omega;\Phi) = \int \frac{d^2k}{(2\pi)^2} 
\left(\frac{\hat{\bf q}\times {\bf k}}{m^*}\right)^2
\frac{f({\cal E}^\alpha_{{\bf k} + {\bf q}}) - f({\cal E}^\beta_{\bf k})}{i\omega - {\cal E}^\alpha_{{\bf k} + {\bf q}} + {\cal E}^\beta_{\bf k}},\nonumber\\
\label{pi11}
\end{eqnarray}
$\Pi^{\alpha\beta}_{10} = \Pi^{\alpha\beta}_{01} = 0$, the indices $\alpha$ and $\beta$ can be either $S$ or $A$, and ${\cal E}^S_{\bf k}$ and ${\cal E}^A_{\bf k}$ are the shifted energy dispersions given in (\ref{dispS}) and (\ref{dispA}).  At $T=0$ the integrals (\ref{pi00}) and (\ref{pi11}) can be performed analytically to give closed-form expressions for ${\cal K}^\pm_{00}$ and ${\cal K}^\pm_{11}$ (see Appendix).  

There is a qualitative change in the $\Phi$ dependence of ${\cal K}_{00}^\pm$ and ${\cal K}_{11}^\pm$ when $\Phi = E_F$ (here, as in Sec.~\ref{secII}, $E_F = k_F^2/(2m^*)$ is the Fermi energy when $\Phi = 0$).  As noted in Sec.~\ref{secII}, for $\Phi < E_F$ there are two Fermi surfaces and for $\Phi > E_F$ there is one Fermi surface.  In the latter case the antisymmetric band is empty and, at $T=0$, the Fermi function $f(E_k^A) = 0$ for all ${\bf k}$.  Thus, while the out-of-phase response functions continue to evolve with $\Phi$ for $\Phi > E_F$ due to virtual transitions from the symmetric band to the antisymmetric band, the in-phase response functions, which only involve intraband transitions, become $\Phi$ independent for $\Phi > E_F$.

Using the RPA action (\ref{RPA_action}) we can study the effect that introducing the order parameter $\Phi$ has on the collective modes of the system.  These modes naturally decouple into in-phase and out-of-phase sectors and their dispersions are determined by the poles of the gauge field propagators after analytic continuation to the real frequency axis.  These poles occur when the determinant of the inverse of the matrix formed by the analytically continued gauge field propagators, ${D^{\pm}}^{-1} ({\bf q},\omega;\Phi) \equiv {\cal D^{\pm}}^{-1}({\bf q},i\omega \rightarrow \omega + i\epsilon;\Phi)$, vanishes, and are thus obtained by solving the equation,
\begin{eqnarray}
\det {D^{\pm}}^{-1} = K_{00}^\pm({\bf q},\omega;\Phi) K_{11}^\pm({\bf q},\omega;\Phi) - \frac{q^2}{(2\pi\lambda)^2} = 0,\nonumber\\
\label{collective}
\end{eqnarray}
in the in-phase $(+)$ and out-of-phase $(-)$ sectors. Here $K_{00}^\pm({\bf q},\omega;\Phi) = {\cal K}_{00}^\pm({\bf q},i\omega \rightarrow \omega+i\epsilon;\Phi)$ and $K_{11}^\pm({\bf q},\omega;\Phi) = {\cal K}_{11}^\pm({\bf q},i\omega \rightarrow \omega+i\epsilon;\Phi)$ are, respectively, the bare density and transverse-current response functions in these two sectors.

When $\Phi = 0$, the two layers are decoupled and the bare response functions, and hence the collective mode dispersions, are the same in the in-phase and the out-of-phase sectors. In the limit $\omega \gg v_F q$, these response functions are given approximately by
\begin{eqnarray}
K_{00}^\pm({\bf q},\omega;0) &\simeq& - \frac{E_F}{2\pi} \frac{q^2}{\omega^2},\label{k00+}\\ 
K_{11}^\pm({\bf q},\omega;0) &\simeq& -\frac{E_F}{2\pi},\label{k11+}
\end{eqnarray}
which can be expressed in a more familiar form using the fact that $E_F/(2\pi) = n/m^*$. The solution to (\ref{collective}) in the $q\rightarrow 0$ limit then yields modes with frequency, 
\begin{eqnarray} 
\omega^{\pm} &\simeq& \lambda E_F,\label{cyclotron}
\end{eqnarray}
in the in-phase $(\omega^+)$ and out-of-phase $(\omega^-)$ sectors.  Note that $\lambda E_F = e B/m^*$ is the cyclotron frequency for particles of mass $m^*$, consistent with with Kohn's theorem,\cite{kohn61,kohn_note} and indicating that these modes are the $q \rightarrow 0$ in-phase and out-of-phase cyclotron modes.\cite{halperin93} 

\begin{figure*}
  \begin{center}
\includegraphics[width=2\columnwidth]{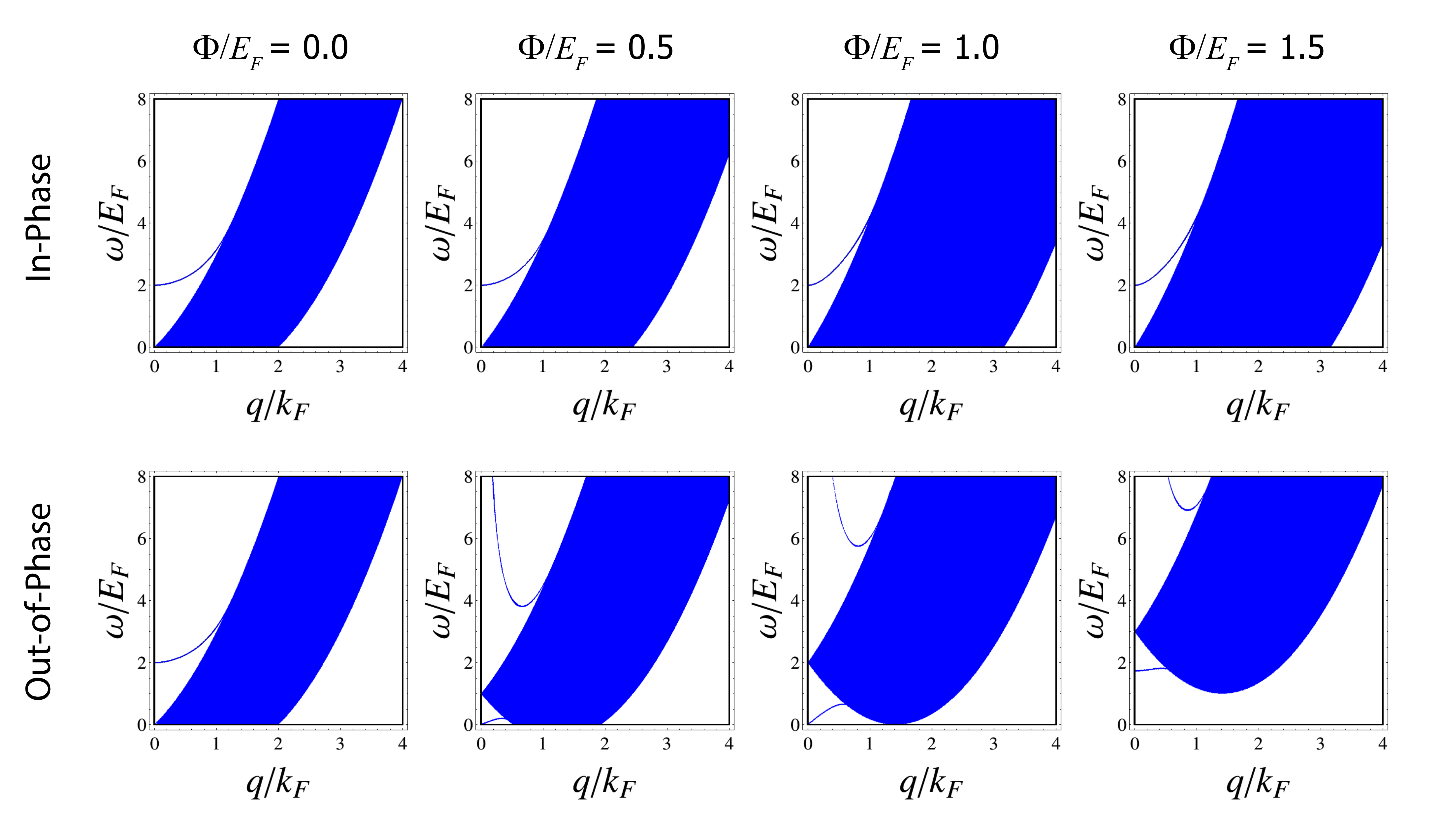}
  \caption{(Color online). Regions in $q, \omega$ space where the imaginary parts of the RPA gauge field propagators (after analytic continuation to the real frequency axis) are nonzero, showing the particle-hole continuum and collective mode excitations in the in-phase (top row) and out-of-phase (bottom row) sectors for different values of the order parameter $\Phi$. For $\Phi = 0$, the excitation spectra are identical in both sectors and show the usual particle-hole continuum and the $q\rightarrow 0$ cyclotron mode.  For $\Phi = 0.5 E_F$, in the in-phase sector the spectrum is only slightly changed (with the particle-hole continuum broadening in $q$ due to the growing symmetric Fermi surface), while in the out-of-phase sector the particle-hole continuum is significantly modified, and both a diverging and a gapless collective mode can be seen as $q\rightarrow 0$. For $\Phi = E_F$, the point at which the antisymmetric Fermi surface vanishes, in the in-phase sector the spectrum is again only slightly changed, and in the out-of-phase sector the low-energy collective mode is still gapless and the particle-hole continuum touches the $\omega = 0$ axis at the point $q = \sqrt{2} k_F$.  For $\Phi = 1.5 E_F$, in the in-phase sector the spectrum is identical to the case $\Phi = E_F$ (as it is for all $\Phi > E_F$), and in the out-of-phase sector the particle-hole continuum and low-energy collective mode are now fully gapped (as they are for all $\Phi > E_F$).  Results are for $\lambda = 2$. } \label{ph_continuum}
   \end{center}
\end{figure*}

When $\Phi \ne 0$, in the $\omega \gg v_F q$ limit the leading contributions to the in-phase response functions given above are unchanged.  As a consequence, the energy of the in-phase cyclotron mode at $q=0$ is also unchanged, again consistent with Kohn's theorem, although the leading $O(q^2)$ contribution to the dispersion (obtained by solving (\ref{collective}) using the expressions given in the Appendix for $K_{00}^+$ and $K_{11}^+$ which include the $O(q^4)$ and $O(q^2)$ contributions, respectively) is modified as follows,
\begin{eqnarray}
{\omega^+}^2 \simeq \left\{
\begin{array}{cc}\lambda^2 E_F^2 + 4 (E_F^2 + \Phi^2) \frac{q^2}{k_F^2}, & ~\Phi < E_F,\\
\lambda^2 E_F^2 + 8 E_F^2 \frac{q^2}{k_F^2}, & ~\Phi > E_F.
\end{array}\right.
\label{wp}
\end{eqnarray}
The out-of-phase response functions, however, are significantly altered.  In the $|\omega - 2\Phi| \gg v_F q$  limit we find for $\Phi < E_F$,
\begin{eqnarray}
K_{00}^-({\bf q},\omega;\Phi<E_F) &\simeq&  \frac{2m^*}{\pi} \frac{\Phi^2}{4\Phi^2-\omega^2},\\
K_{11}^-({\bf q},\omega;\Phi<E_F) &\simeq& \frac{E_F}{2\pi} \frac{\omega^2}{4\Phi^2 -\omega^2},
\label{krm1}
\end{eqnarray}
and for $\Phi > E_F$,
\begin{eqnarray}
K_{00}^-({\bf q},\omega;\Phi>E_F) &\simeq&  \frac{2m^*}{\pi} \frac{E_F\Phi}{4\Phi^2-\omega^2},\label{krm1}\\
K_{11}^-({\bf q},\omega;\Phi>E_F) &\simeq& \frac{E_F}{2\pi} \frac{\omega^2 - 4\Phi(\Phi - E_F) }{4\Phi^2 -\omega^2}.
\label{krm2}
\end{eqnarray}
The long wavelength pole in these response functions at $\omega = 2\Phi$ corresponds to the $q\rightarrow 0$ interband transition from the symmetric band to the antisymmertic band. Using these response functions (including $O(q^2)$ contributions omitted above but given in the Appendix) to solve (\ref{collective}) we find {\it two} out-of-phase collective modes in the long wavelength limit, one low-energy mode, and one high-energy mode. 

For $\Phi <  E_F$ the low-energy mode is gapless with linear dispersion,
\begin{eqnarray}
\omega^-_1 \simeq \left(\frac{2}{3} \Phi^2 + 8 \frac{\Phi^2}{\lambda^2}\right)^{1/2} \frac{q}{k_F}.\label{wm1}
\end{eqnarray}
This mode couples to the composite fermions as an effective gapless out-of-phase photon.  Even when the order parameter $\Phi$ is finite, provided it is less than $E_F$ in magnitude, this mode remains gapless.  This is due to the fact that the $q \rightarrow 0$ limit of the bare out-of-phase static transverse current response function is $\lim_{q\rightarrow 0} K^-_{11}({\bf q},\omega=0;\Phi< E_F) = 0$.  Thus there is no out-of-phase Meissner effect for composite fermions when $\Phi < E_F$.  This in turn implies the system is compressible to out-of-phase density perturbations (which appear to composite fermions as an out-of-phase magnetic field).  This lack of an out-of-phase Meissner effect for $\Phi < E_F$ can be traced back to the flat density of states, which, as noted above, is also the reason the Stoner energy density $E_S(\Phi)$ is a purely quadratic function of $\Phi$ for $\Phi < E_F$.  

For $\Phi > E_F$, there {\it is} an out-of-phase Meissner effect for composite fermions, with $\lim_{q\rightarrow 0} K_{11}({\bf q},\omega=0;\Phi > E_F) =  -\frac{E_F}{2\pi \Phi} (\Phi - E_F)$.  This leads to a gap opening up in the out-of-phase photon dispersion.  For this dispersion we find
\begin{eqnarray}
\omega^-_1 \simeq \left(\Delta_{q=0}^2 + \left(2 E_F^2 - \frac{4}{3} \Phi E_F +8\frac{\Phi^2}{\lambda^2} \right)\frac{q^2}{k_F^2}\right)^{1/2},\label{wo1lp}
\end{eqnarray}
where the $q=0$ energy gap is
\begin{eqnarray}
\Delta_{q=0} = 2 (\Phi(\Phi - E_F))^{1/2}.\label{deltaq0}
\end{eqnarray}
We note that the transition to the interlayer coherent state is always to a state with $\Phi \ge E_F$ (both at the Stoner level, for which $\Phi$ jumps to $E_F$ at the transition as described in Sec.~\ref{secII}, and when gauge fluctuations are included, for which $\Phi$ jumps to a value larger than $E_F$, see below).  The transition is therefore always to a state which is incompressible in the out-of-phase sector, and thus behaves like a quantum Hall state in the counterflow channel.  This, together with compressibility in the in-phase sector, is the hallmark of the interlayer coherent composite fermion state.\cite{alicea09}

For $\Phi \ne 0$ the layers are coupled and Kohn's theorem no longer holds for the out-of-phase cyclotron mode.  For both $\Phi < E_F$ and $\Phi > E_F$ we find the dispersion of this mode diverges as $q\rightarrow 0$, with
\begin{eqnarray}
{\omega^-_2}^2 \simeq \left\{\begin{array}{cl} 2 \lambda^2 \Phi^2 \frac{k_F^2}{q^2} + \lambda^2 E_F^2 + 8 \Phi^2, & \Phi < E_F,\\
2\lambda^2\Phi E_F\frac{k_F^2}{q^2} + \lambda^2 E_F^2 + 4 \Phi^2 + 4 \Phi E_F, & \Phi > E_F.\end{array}\right. \nonumber \\
\label{wo2}
\end{eqnarray}

The collective modes described above, along with the continuum of particle-hole excitations in the in-phase and out-of-phase sectors, are illustrated in Fig.~\ref{ph_continuum}.  This figure shows the regions in $q$ and $\omega$ space where the analytically continued gauge field propagators $D^\pm_{\mu\nu}({\bf q},\omega;\Phi) = {\cal D}^\pm_{\mu\nu}({\bf q},i\omega \rightarrow \omega + i \epsilon; \Phi)$, evaluated for the case $\lambda = 2$, have nonzero imaginary part for different values of $\Phi$.   For $\Phi = 0$ the layers are decoupled and the excitations are identical in the two sectors, consisting of the usual particle-hole continuum and the cyclotron mode. For $0 < \Phi < E_F$, in the in-phase sector the particle-hole continuum, which consists entirely of intraband excitations, grows broader in $q$ due to the increasing size of the symmetric Fermi surface, but is otherwise only mildly affected, and the cyclotron mode is likewise only slightly modified.   By contrast, in the out-of-phase sector the particle-hole continuum, which consists entirely of interband excitations, is altered significantly and both the diverging out-of-phase cyclotron mode and gapless out-of-phase ``photon" mode described above can be seen.  For $\Phi > E_F$ the in-phase excitations are independent of $\Phi$ (due to the fact that there is only one Fermi surface) while the out-of-phase excitations continue to evolve, with gaps appearing both in the interband particle-hole continuum at $q = \sqrt{2} k_F$ ($\Delta_{q=\sqrt{2} k_F}$, see (\ref{deltaqs2})) and the out-of-phase photon mode at $q=0$ ($\Delta_{q=0}$, see (\ref{deltaq0})).  

It is apparent that the order parameter $\Phi$ has a much stronger effect on the out-of-phase gauge propagators than on the in-phase gauge propagators.  This is consistent with our expectation that it is the out-of-phase gauge fluctuations which strongly suppress the formation of the interlayer coherent composite fermion state.  To analyze this suppression we use an approach introduced by Ubbens and Lee\cite{ubbens94_1} to study BCS pairing of spinons in an effective gauge-theory description of the $t$-$J$ model.  In this approach, we calculate the gauge fluctuation contribution to the correlation energy within the RPA in the presence of the order parameter $\Phi$.  While this calculation does not go beyond mean-field theory in $\Phi$, which we continue to assume is constant in time and independent of position, it does go beyond the composite fermion mean-field theory by including gaussian fluctuations of the gauge fields.

Integrating out the gauge fields in (\ref{RPA_action}) and taking the $T \rightarrow 0$ limit of the free energy we obtain the following contribution to the energy density of the in-phase and out-of-phase gauge fluctuations, 
\begin{eqnarray}
E_{CS}^{\pm}(\Phi) &=& \frac{1}{2} \int_{-\infty}^\infty \frac{d\omega}{2\pi} \int \frac{d^2 q}{(2\pi)^2} \ln \det {{\cal D}^\pm}^{-1}({\bf q},i\omega;\Phi).\nonumber\\
\label{ECS}
\end{eqnarray}
The change in energy density due to introducing the order parameter, $\Delta E_{CS}^{\pm}(\Phi) = E_{CS}^{\pm}(\Phi) - E_{CS}^\pm(0)$, can then be expressed as the following integral over dimensionless variables $\bar q = q/k_F$ and $\bar \omega = \omega/E_F$,
\begin{eqnarray}
\frac{\Delta E_{CS}^{\pm}(\Phi)}{\nu_0 E_F^2} &=& \frac{1}{\pi}\int_{0}^\infty d\bar\omega \int_0^\infty \bar q d \bar q \nonumber \\
&&\ln\frac{\bar q^2-(2\pi \lambda)^2 \bar {\cal K}^{\pm}_{00}(\bar q,\bar \omega;\bar \Phi) \bar {\cal K}^{\pm}_{11}(\bar q,\bar\omega;\bar\Phi)}{\bar q^2-(2\pi \lambda)^2 \bar {\cal K}^{\pm}_{00}(\bar q,\bar\omega;0) \bar {\cal K}^{\pm}_{11}(\bar q,\bar\omega;0)},\nonumber\\
\label{rpa_integral}
\end{eqnarray} 
where $\bar\Phi = \Phi/E_F$ and, as in Sec.~\ref{secII}, $\nu_0 = m^*/(2\pi)$ is the density of states per layer and $E_F = k_F^2/(2m^*)$ is the Fermi energy for $\Phi = 0$. Here we have used the fact that ${\cal K}_{00}^\pm$ and ${\cal K}_{11}^\pm$ can be expressed as
\begin{eqnarray}
{\cal K}_{00}^{\pm}({\bf q},i\omega;\Phi) &=& m^* {\bar {\cal K}}_{00}^{\pm}(\bar q,i\bar\omega,\bar \Phi),\label{k00pmbar}\\
{\cal K}_{11}^{\pm}({\bf q},i\omega;\Phi) &=& \frac{k_F^2}{m^*} {\bar {\cal K}}_{11}^{\pm}(\bar q,i\bar\omega;\bar\Phi),\label{k11pmbar}
\end{eqnarray}
where $\bar {\cal K}_{00}^{\pm}$ and $\bar {\cal K}_{11}^\pm$ are dimensionless functions of $\bar q$, $\bar\omega$, and $\bar\Phi$.  Using the analytic expressions for ${\cal K}^\pm_{00}$ and ${\cal K}^\pm_{11}$ given in the Appendix, we need only numerically perform a single two-dimensional integral to determine $\Delta E^+_{CS}(\Phi)$ or $\Delta E^-_{CS}(\Phi)$ for a given value of $\Phi$ and $\lambda$.

Before presenting the results of this full integration,  it is instructive to analyze the behavior of $\Delta E^+_{CS}(\Phi)$ and $\Delta E^-_{CS}(\Phi)$ in the $\Phi \rightarrow 0$ limit.   In both cases the integrand in (\ref{rpa_integral}) can be Taylor expanded to second order in $\Phi$ using our analytic expressions for ${\cal K}^\pm_{00}$ and ${\cal K}^\pm_{11}$.  For $\Delta E_{CS}^+(\Phi)$ the integral over $q$ and $\omega$ can then be carried out to yield a finite coefficient of the $O(\Phi^2)$ contribution.  Performing this integration numerically for $\lambda = 2$ we find
\begin{eqnarray}
\frac{\Delta E_{CS}^+(\Phi)}{\nu_0 E_F^2} \simeq - 0.57 \frac{\Phi^2}{E_F^2}.\label{DECSP}
\end{eqnarray}
Thus the in-phase gauge fluctuations contribute a term to the total energy which is analytic in $\Phi$ and, because it is negative, favors the formation of the interlayer coherent composite fermion state.  While it is not possible to analytically determine the $\lambda$ dependence of $\Delta E_{CS}^+(\Phi)$, even in the small $\Phi$ limit, if we expand the integrand in (\ref{rpa_integral}) to second order in both $\lambda$ and $\Phi$ and carry out the integration we find that $\Delta E_{CS}^+(\Phi)/(\nu_0 E_F^2) \simeq -0.14 \lambda^2 \Phi^2$ for small $\lambda$.  The fact that the magnitude of this contribution grows with increasing $\lambda$ is consistent with $\lambda$ being a measure of the strength of the gauge fluctuations in the system. 

By contrast,  when the integrand in (\ref{rpa_integral}) for $\Delta E_{CS}^-(\Phi)$ is expanded to second order in $\Phi$ and integrated over $q$ and $\omega$ the coefficient of the $O(\Phi^2)$ term diverges, indicating that $\Delta E_{CS}^-(\Phi)$ is not analytic in $\Phi$.  We find that this divergence arises from the $\bar q \ll 1$ region of the $q,\omega$ integration.  The leading nonanalytic behavior in $\Delta E_{CS}^-(\Phi)$ can then be isolated by expressing the integral (\ref{rpa_integral}) as a sum of two integrals, one where $q$ is integrated from 0 to a cutoff $q_c$ and a second where $q$ is integrated from $q_c$ to infinity. Regardless of the value of the cutoff $q_c$ the second integral will be analytic in $\Phi$ and contribute a term of $O(\Phi^2)$ to the energy. The leading nonanalytic behavior of $\Delta E_{CS}^-(\Phi)$ for small $\Phi$ is thus contained in the first integral.  For this integral, rather than expanding the integrand, we can expand the argument of the logarithm in the integrand, first to second order in $\Phi$ and then in powers of $q$.  After doing so, using the expressions for ${\cal K}_{00}^-$ and ${\cal K}_{11}^-$ from the Appendix, we find,
\begin{eqnarray}
\frac{\Delta E^{-}_{CS}(\Phi)}{\nu_0 E_F^2}
&\simeq& \frac{1}{\pi} \int_0^\infty d\bar\omega \int_0^{q_c} \bar q d\bar q  
\ln\left(1+ \frac{2\lambda^2}{\lambda^2 + {\bar \omega}^2} \frac{\bar\Phi^2}{{\bar q}^2}\right).\nonumber\\ \label{deltaecs_approx}
\end{eqnarray}

\begin{figure}
\begin{center}
\includegraphics[width=\columnwidth]{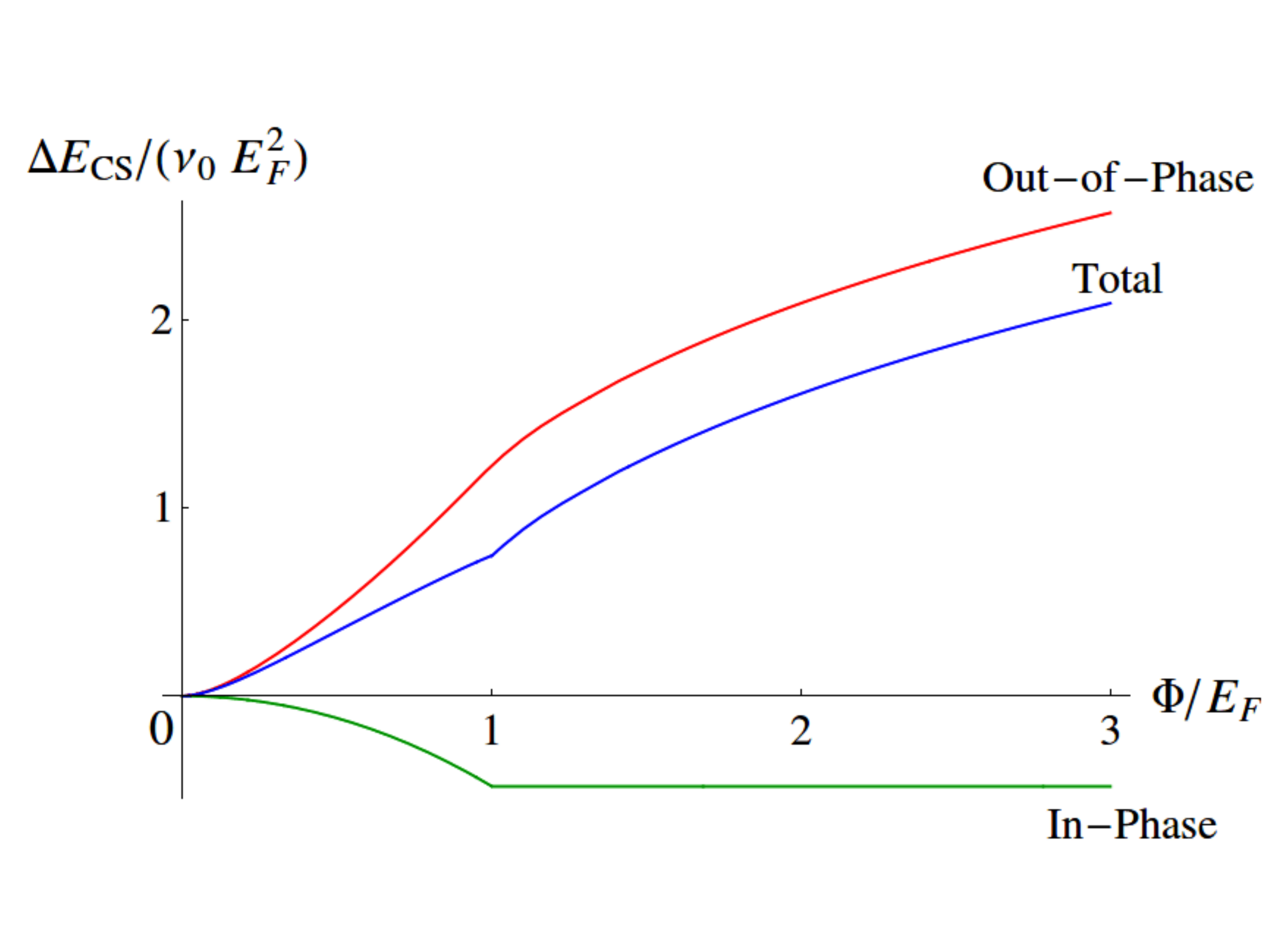}
\caption{(Color online). RPA contribution to the correlation energy density from in-phase (green) and out-of-phase (red) gauge fluctuations, and their total (blue) as a function of the order parameter $\Phi$.  Results are for $\lambda = 2$.} \label{delta_Ecs}
   \end{center}
\end{figure}

The $\bar \omega$ integration in (\ref{deltaecs_approx}) can be performed to obtain
\begin{eqnarray}
\frac{\Delta E^{-}_{CS}(\Phi)}
{\nu_0 E_F^2} &\simeq& \int_0^{q_c}  {\bar q} d {\bar q} \left(\left(\frac{2 \lambda^2 {\bar \Phi}^2}{{\bar q}^2} + \lambda^2 \right)^{1/2} - \lambda\right).\label{zpe} 
\end{eqnarray}
This integral has a clear physical meaning; it is the difference in the zero-point energies associated with the out-of-phase cyclotron mode $\omega_2^-$ for the case $\Phi \ne 0$ (which diverges as $q \rightarrow 0$) and $\Phi = 0$ (which remains finite as $q\rightarrow 0$).  The  singular contribution to (\ref{zpe}) can be found by carrying out the $\bar q$ integration to leading logarithmic accuracy with the result
\begin{eqnarray}
\frac{\Delta E^{-}_{CS}(\Phi)}{\nu_0 E_F^2} &\simeq& \lambda~ \frac{\Phi^2}{E_F^2} \left|\ln \frac{\Phi}{E_F}\right|,\label{DECSM}
\end{eqnarray}
which is asymptotically exact in the $\Phi \rightarrow 0$ limit for all values of $\lambda$. Because it is singular, this positive energy cost for introducing a nonzero $\Phi$ will always dominate the total energy of the system for small enough $\Phi$, regardless of the value of the coupling constant $g$.  This reflects the fact that the out-of-phase gauge fluctuations strongly inhibit the formation of the interlayer coherent composite fermion state.  Note that, like $\Delta E^{+}_{CS}(\Phi)$, $\Delta E^{-}_{CS}(\Phi)$ grows in magnitude with increasing $\lambda$, again consistent with $\lambda$ being a measure of the strength of the gauge fluctuations in the system.

Results for numerically performing the full integral (\ref{rpa_integral}) for the case $\lambda = 2$ are shown in Fig.~\ref{delta_Ecs}.  This plot shows the dependence of the in-phase, $\Delta E_{CS}^+(\Phi)$, and out-of-phase, $\Delta E_{CS}^-(\Phi)$, contributions to the energy on $\Phi$, as well as their sum, $\Delta E_{CS}(\Phi) = \Delta E^+_{CS}(\Phi) + \Delta E^-_{CS}(\Phi)$.  For $\Phi < E_F$ the in-phase contribution is negative and decreases with increasing $\Phi$, consistent with the small $\Phi$ behavior found above, and confirming that this contribution favors the formation of an interlayer coherent composite fermion state.  The out-of-phase contribution is significantly larger in magnitude than the in-phase contribution and increases with increasing $\Phi$, indicating that this contribution strongly suppresses the formation of the interlayer coherent composite fermion state, again consistent with the above small $\Phi$ analysis.   Note that for $\Phi > E_F$ the in-phase contribution becomes independent of $\Phi$, because the in-phase response functions do not change once $\Phi$ exceeds $E_F$, while the out-of-phase contribution continues to grow. Thus, for all values of $\Phi$, the out-of-phase contribution dominates and the total gauge field contribution to the energy density, $\Delta E_{CS} (\Phi)$, grows monotonically with increasing $\Phi$.

\begin{figure}
\begin{center}
\includegraphics[width=\columnwidth]{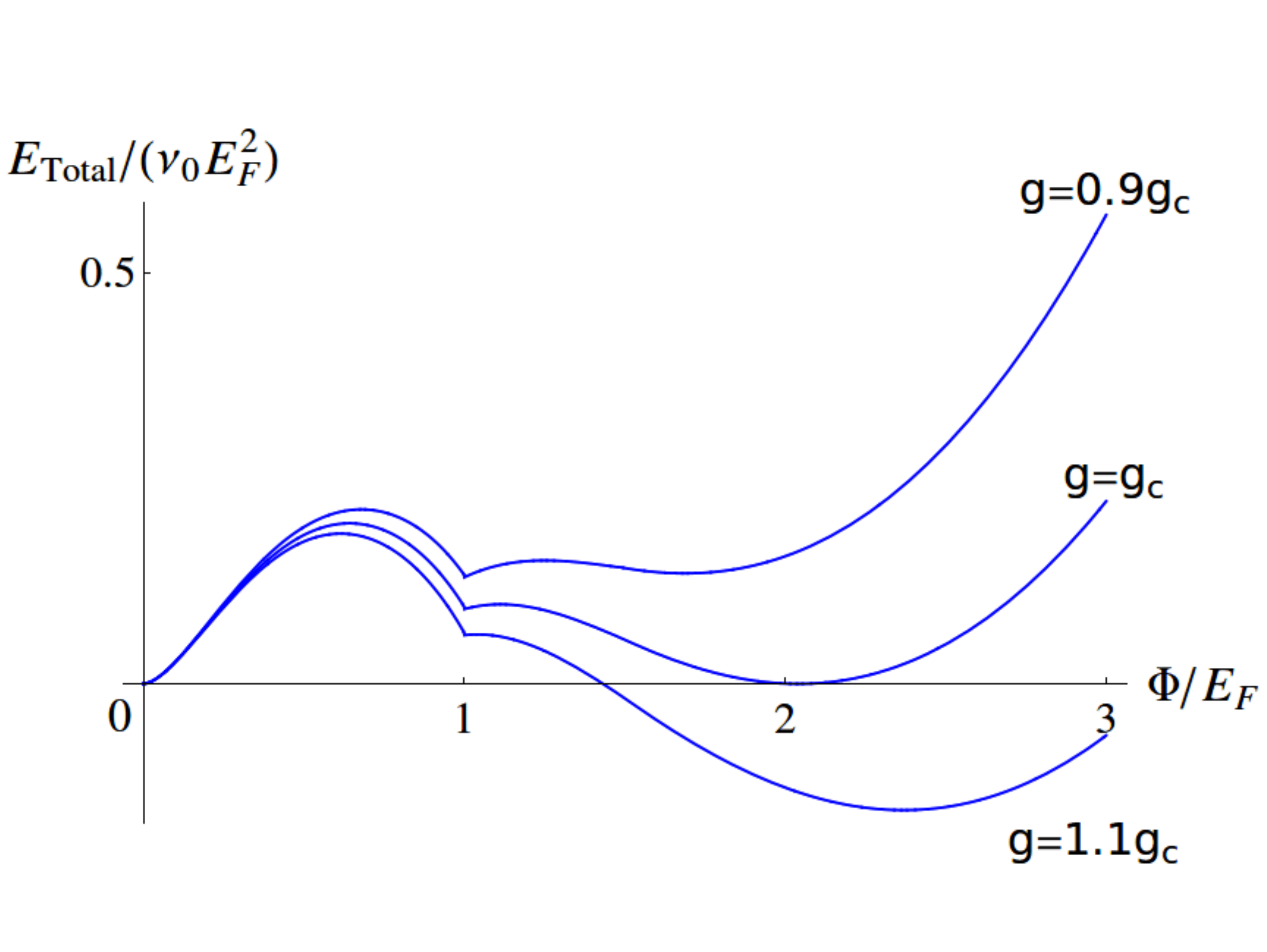}
\caption{(Color online). Total energy density obtained by adding the Stoner $(E_S(\Phi))$ and gauge field $(\Delta E_{CS}(\Phi))$ contributions, plotted as a function of the order parameter $\Phi$ for coupling strengths $g=0.9 g_c, g_c$ and $1.1 g_c$ where $g_c \simeq 2.9$ is the critical value of the coupling constant.  Results are for $\lambda = 2$.} \label{etotal}
\end{center}
\end{figure}

The total energy density for the system is obtained by adding the RPA gauge fluctuation contribution to the Stoner energy found in Sec.~\ref{secII} to give $E_{\rm Total}(\Phi) = E_{S}(\Phi) + \Delta E_{CS}(\Phi)$.  Figure \ref{etotal} shows $E_{\rm Total}(\Phi)$ plotted as a function of $\Phi$ for different values of the dimensionless coupling constant $g$ as the system undergoes a first-order phase transition from decoupled bilayers ($\Phi = 0$) to the interlayer coherent composite fermion state ($\Phi \ne 0$) for the case $\lambda = 2$.  For a given $g$ the order parameter is found by minimizing the energy as a function of $\Phi$.  When gauge fluctuations are included, as the coupling constant $g$ is increased from 0, the energy is minimized by a nonzero $\Phi$ at the critical value $g = g_c \simeq 2.9$, which should be compared to the critical value $g = 1$ for the Stoner analysis when gauge fluctuations are ignored (see Fig.~\ref{stoner}).  If we assume the relation $g \simeq 3 l_0/d$ holds this implies that the gauge fluctuations have shifted the critical layer spacing from $(d/l_0)_c\sim 3$ down to $(d/l_0)_c\sim 1$ which, we note, is below the critical layer spacing for the $\nu_{tot}=1$ bilayer quantum Hall state,  theoretical estimates of which range from $d/l_0 \simeq 1.3$ (Refs.~\onlinecite{moon95,schliemann01}) to $d/l_0 \simeq 1.6$ (Ref.~\onlinecite{shibata06}). This shifting down of $(d/l_0)_c$ may account for the fact that the interlayer coherent composite fermion state has not yet been observed experimentally.

\begin{figure}
\begin{center}
\includegraphics[width=\columnwidth]{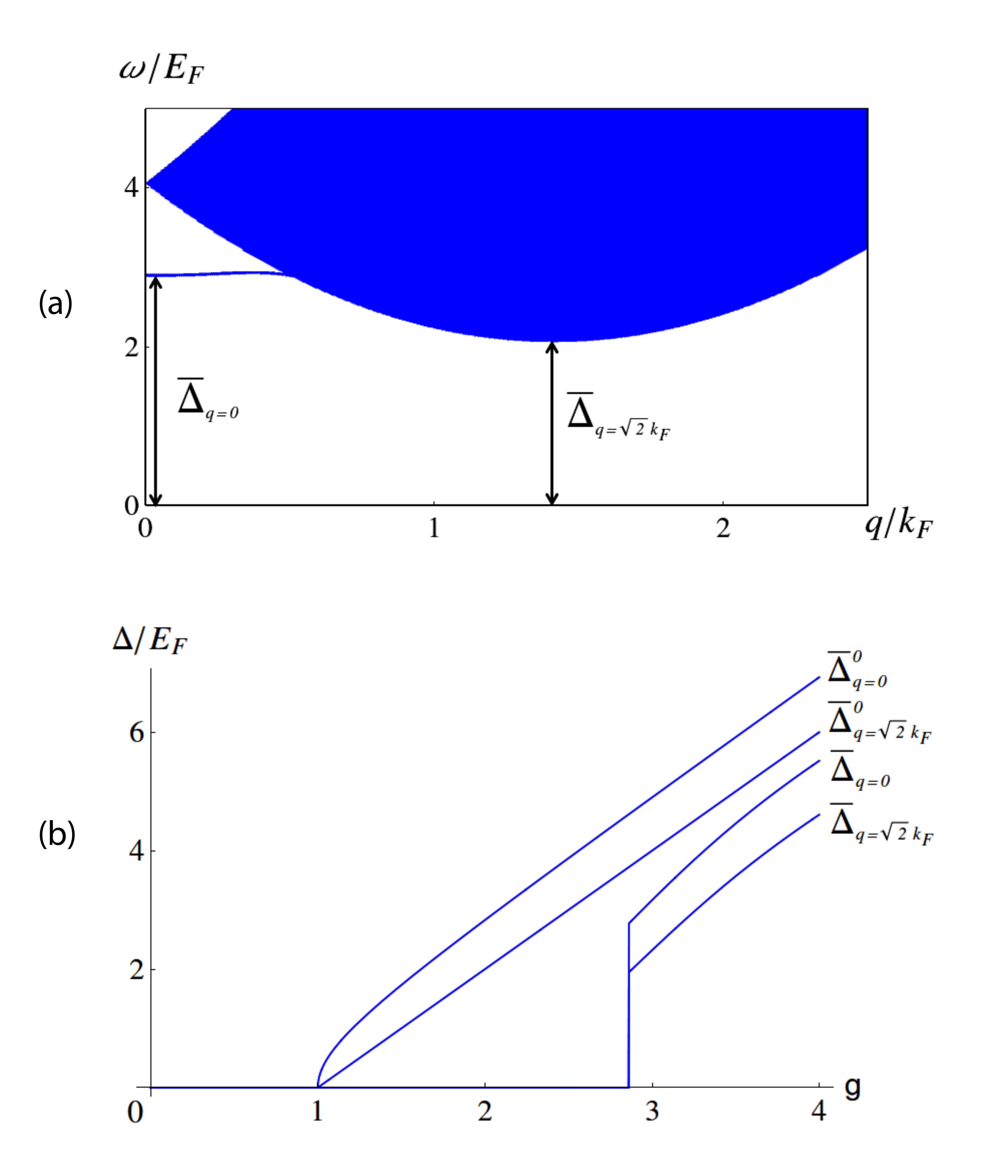}
\caption{(Color online).
(a) $q$,$\omega$ plot showing the energy dispersion of the low-energy collective mode and particle-hole continuum in the out-of-phase sector immediately after the transition into the interlayer coherent composite fermion state at $g = g_c \simeq 2.9$ where $\Phi \simeq 2.1 E_F$. The energy gaps in the collective mode at $q=0$ ($\bar \Delta_{q=0} = \Delta_{q=0}/E_F$) and in the particle-hole spectrum at $q=\sqrt{2} k_F$ ($\bar \Delta_{q=\sqrt{2} k_F} = \Delta_{q=\sqrt{2} k_F}/E_F$) are indicated. (b) Energy gaps $\Delta_{q=0}$ and $\Delta_{q=\sqrt{2} k_F}$ as a function of the coupling constant $g$.  Results are shown both for the simple Stoner analysis of Sec.~\ref{secII} where the gaps (with superscript 0) open continuously at the transition, and for when the gauge fluctuation contribution to the energy is included where the gaps (without superscript 0) jump discontinuously at the transition.  Results are for $\lambda = 2$.} \label{gaps}
\end{center}
\end{figure}

In addition to increasing the coupling strength required to produce the transition to the interlayer coherent composite fermion state, the gauge fluctuations lead to a qualitative change in the nature of this transition.   This change is seen in the dependence of the energy gaps in the out-of-phase sector when the transition occurs, both for the interband particle-hole excitations at $q = \sqrt{2} k_F$ ($\Delta_{q=\sqrt{2} k_F} \propto (\Phi - E_F)$ for $\Phi -E_F>0$, see (\ref{deltaqs2})) and for the long wavelength out-of-phase photon mode at $q=0$ ($\Delta_{q=0} \propto (\Phi - E_F)^{1/2}$ for small $\Phi -E_F >0$, see (\ref{deltaq0})).  As shown in Sec.~\ref{secII}, when gauge fluctuations are ignored the value the order parameter takes immediately after the transition at $g=1$ is $\Phi = E_F$.  The order parameter then grows continuously for $g>1$ and the out-of-phase energy gaps open continuously.   When gauge fluctuations are included, not only does the critical coupling constant increase from $g=1$ to $g_c \simeq 2.9$ for $\lambda = 2$, but the value the order parameter takes immediately after the transition occurs increases from $\Phi = E_F$ to $\Phi \simeq 2.1 E_F$.  Figure \ref{gaps}(a) shows the excitation spectrum in the out-of-phase sector for $\Phi \simeq 2.1 E_F$.  Because $\Phi > E_F$ this spectrum is fully gapped, both at $q=0$ and $q=\sqrt{2} k_F$.  Thus we see there is a discontinuous jump in $\Delta_{q=0}$ and $\Delta_{q=\sqrt{2} k_F}$ at the transition when gauge fluctuations are included.  Figure \ref{gaps}(b) shows plots of $\Delta_{q=0}$ and $\Delta_{q=\sqrt{2} k_F}$ as a function of $g$.  Results are shown both for the case when gauge fluctuations are ignored and the gaps open continuously at the Stoner critical coupling $g=1$, and when gauge fluctuations are included and the gaps jump discontinuously at the increased critical coupling $g_c \simeq 2.9$.  We believe the result that gauge fluctuations lead to a discontinuous jump in the out-of-phase energy gaps at this transition is likely to be valid beyond the level of the RPA calculation presented here.  Thus, if a transition to an interlayer coherent composite fermion state is observed, the measurement of such a jump would provide indirect experimental evidence for the presence of gauge fluctuations in the system.

\section{Conclusions}
\label{secIV}

We have studied the effect of fluctuations in the Chern-Simons gauge fields on the possible formation of the interlayer coherent composite fermion state proposed in Ref.~\onlinecite{alicea09} in a symmetrically doped $\nu_{tot} = 1$ quantum Hall bilayer. Scattering from these gauge fields leads to layer-dependent fluctuations in the Aharonov-Bohm phase experienced by composite fermions as they propagate through the bilayer, strongly suppressing any interlayer phase coherence these composite fermions may have.  This suppression manifests itself through the appearance of a contribution to the ground state energy from gauge fluctuations which is logarithmically singular in the order parameter characterizing interlayer coherence, and which grows monotonically as this order parameter increases from zero.

If the gauge field contribution to the energy is ignored, the transition from two decoupled single-layer composite fermion metals to an interlayer coherent composite fermion state with increasing interlayer coupling is a simple Stoner instability, and the energy gaps to out-of-phase excitations open continuously from zero at the transition.  When the gauge field contribution to the energy is included there are two main effects: (1) the interlayer coupling strength required to drive the transition grows substantially (contrast Fig.~\ref{etotal} with Fig.~\ref{stoner}); and (2) the out-of-phase energy gaps jump discontinuously from zero to a finite value at the transition (see Fig.~\ref{gaps}).   The first effect may account for the fact that the interlayer coherent state has not yet been observed experimentally in $\nu_{tot}=1$ bilayers.  The second effect suggests that if such a transition were to be observed, the detection of a discontinuous jump in the out-of-phase energy gaps would provide indirect experimental evidence for the presence of gauge fluctuations in the system.  Of more general interest, we believe that the model studied here provides a novel example of the qualitative effects that gauge fluctuations can have on quantum phase transitions in dense Fermi systems. 

\acknowledgments

The authors thank Yafis Barlas and Yong Baek Kim for helpful discussions.  This work was supported by US DOE Grant DE-FG02-97ER45639.

\onecolumngrid

\appendix*

\section{Calculation of ${\cal K}_{00}^\pm$ and ${\cal K}_{11}^\pm$}

To determine ${\cal K}_{00}^\pm$ and ${\cal K}_{11}^\pm$ using (\ref{kpm}) we need to evaluate $\Pi_{00}^\pm$ and $\Pi_{11}^\pm$ which are defined in (\ref{pi+}) and (\ref{pi-}) in terms of the integrals (\ref{pi00}) and (\ref{pi11}).  Using these expressions we find that
\begin{eqnarray}
\Pi_{00}^{+}({\bf q},i\omega;\Phi) &=& -\frac{1}{2}\Bigr(F_1({\bf q},i\omega;k_F^S)+F_1({\bf q},-i\omega;k_F^A)
+F_1({\bf q},i\omega;k_F^A)+F_1({\bf q},-i\omega;k_F^S)\Bigl),\\
\Pi_{11}^{+}({\bf q},i\omega;\Phi) &=& -\frac{1}{2}\Bigl(F_2({\bf q},i\omega;k_F^S)+F_2({\bf q},-i\omega;k_F^A)
+F_2({\bf q},i\omega;k_F^A)+F_2({\bf q},-i\omega;k_F^S)\Bigr),\\
\Pi_{00}^{-}({\bf q},i\omega;\Phi) &=& -\frac{1}{2}\Bigl(F_1({\bf q},i\omega-2\Phi;k_F^S)+F_1({\bf q},-i\omega+2\Phi;k_F^A)
+F_1({\bf q},i\omega+2\Phi;k_F^A)+F_1({\bf q},-i\omega-2\Phi;k_F^S)\Bigr),\\
\Pi_{11}^{-}({\bf q},i\omega;\Phi) &=& -\frac{1}{2}\Bigl(F_2({\bf q},i\omega-2\Phi;k_F^S)+F_2({\bf q},-i\omega+2\Phi;k_F^A)
+F_2({\bf q},i\omega+2\Phi;k_F^A)+F_2({\bf q},-i\omega-2\Phi;k_F^S)\Bigr).
\end{eqnarray}
Here $k_F^S$ and $k_F^A$ are given by (\ref{kfS}) and (\ref{kfA}) for $\Phi < E_F$ and $k_F^S = \sqrt{2} k_F$ and $k_F^A = 0$ for $\Phi > E_F$.  As in the main text, $k_F$ and $E_F = k_F^2/(2m^*)$  are the Fermi wavevector and Fermi energy for $\Phi = 0$, respectively.  The functions $F_1$ and $F_2$ are given by
\begin{eqnarray}
F_1({\bf q},\gamma;k^\alpha_F) = \int_{|{\bf k}| < k^\alpha_F} \frac{d^2 k}{(2\pi)^2} \frac{1}{\gamma- {\cal E}_{{\bf k} + {\bf q}} + {\cal E}_{\bf k}} = m^*
f_1\left(\frac{q}{k_F^\alpha},\frac{1}{2}\left(\frac{\gamma}{E^\alpha}-\frac{q^2}{{k_F^\alpha}^2}\right)\right),
\end{eqnarray}
and 
\begin{eqnarray}
F_2({\bf q},\gamma;k^\alpha_F) = \int_{|{\bf k}| < k_F^\alpha} \left(\frac{\hat{\bf q}\times {\bf k}}{m^*}\right)^2 \frac{d^2 k}{(2\pi)^2} \frac{1}{\alpha - {\cal E}_{{\bf k} + {\bf q}} + {\cal E}_{\bf k}} = \frac{{k_F^\alpha}^2}{m^*} 
f_2\left(\frac{q}{k_F^\alpha},\frac{1}{2}\left(\frac{\gamma}{E^\alpha}-\frac{q^2}{{k_F^\alpha}^2}\right)\right),
\end{eqnarray}
where $E^\alpha = {k^\alpha_F}^2/(2m^*)$; and $f_1$ and $f_2$ are given by the dimensionless integrals
\begin{eqnarray}
f_1(y,z) = \frac{1}{(2\pi)^2}\int_0^1 xdx \int_0^{2\pi} d\theta\frac{1}{z - x  y \cos\theta },\label{f1}
\end{eqnarray}
and
\begin{eqnarray}
f_2(y,z) = \frac{1}{(2\pi)^2}\int_0^1 xdx \int_0^{2\pi} d\theta
\frac{x^2\sin^2\theta}{z - x y \cos\theta }.\label{f2}
\end{eqnarray}
which can be carried out analytically with the results
\begin{eqnarray}
f_1(y,z) = \frac{1}{2\pi y} \frac{z}{y}\left(1-\left(1-\frac{y^2}{z^2}\right)^{1/2}\right),
\label{eq:}
\end{eqnarray}
and
\begin{eqnarray}
f_2(y,z) = \frac{1}{4\pi y} \frac{z}{y}\left(1 - \frac{2}{3} \frac{z^2}{y^2} \left(1-\left(1-\frac{y^2}{z^2}\right)^{3/2}\right)\right),
\end{eqnarray}
where $y$ is real and the branch cuts in the complex $z$-plane of the $(\cdots)^{1/2}$ and $(\cdots)^{3/2}$ expressions are taken along the real axis between the points $z=\pm y$.  

Taken together the above results give closed-form analytic expressions for ${\cal K}_{00}^\pm$ and ${\cal K}_{11}^\pm$ which give, in turn, an analytic expression for the integrand in (\ref{rpa_integral}).  It is then only necessary to carry out a single two-dimensional numerical integral over $q$ and $\omega$ to obtain $\Delta E_{CS}^+$ or $\Delta E_{CS}^-$ for a given value of $\Phi$ and $\lambda$. 

With the branch cuts specified for $f_1$ and $f_2$ it is straightforward to analytically continue ${\cal K}_{00}^\pm$ and ${\cal K}_{11}^\pm$ to the real frequency axis to obtain the bare density and transverse-current response functions: $K_{00}({\bf q},\omega;\Phi) = {\cal K}_{00}({\bf q},i\omega \rightarrow \omega + i\epsilon;\Phi)$ and $K_{11}({\bf q},\omega;\Phi) = {\cal K}_{11}({\bf q},i\omega \rightarrow \omega+ i\epsilon;\Phi)$.  These functions can then be used to find the collective mode dispersions by solving (\ref{collective}).  To find these dispersions, including the $O(q^2)$ terms in $\omega^+$ and $\omega_1^-$, and the $O(q^0)$ term in $\omega_2^-$ we need the following expressions for $K_{00}^\pm$ and $K_{11}^\pm$ valid for small $q$. For the in-phase response functions, when $\Phi < E_F$ and $\omega \gg k_F q/m^*$,
\begin{eqnarray}
K_{00}^+({\bf q},\omega;\Phi < E_F) \simeq -\frac{E_F}{2\pi} \frac{q^2}{\omega^2} - \frac{3(E_F^2 + \Phi^2) }{4 \pi m^*} \frac{q^4}{\omega^4},
\end{eqnarray}
\begin{eqnarray}
K_{11}^+({\bf q},\omega;\Phi < E_F) \simeq - \frac{E_F}{2\pi} - \frac{E_F^2 + \Phi^2}{4\pi m ^*} \frac{q^2}{\omega^2}.
\end{eqnarray}
For $\Phi > E_F$ the value of $\Phi$ in the above expressions is simply replaced by $E_F$,
\begin{eqnarray}
K_{00}^+({\bf q},\omega;\Phi > E_F) \simeq -\frac{E_F}{2\pi} \frac{q^2}{\omega^2} - \frac{3 E_F^2  }{2\pi m^*} \frac{q^4}{\omega^4},
\end{eqnarray}
\begin{eqnarray}
K_{11}^+({\bf q},\omega;\Phi > E_F) \simeq - \frac{E_F}{2\pi} -\frac{E_F^2}{2\pi m^*} \frac{q^2}{\omega^2}.
\end{eqnarray}
For the out-of-phase response functions, when $\Phi < E_F$ and $|\omega - 2\Phi| \gg k_F q /m^*$ we have
\begin{eqnarray}
K_{00}^-({\bf q},\omega;\Phi < E_F) \simeq \frac{2 m^*}{\pi} \frac{\Phi^2}{4\Phi^2 - \omega^2} + 
 \frac{E_F}{2\pi}\frac{(12 \Phi^2 \omega^2 + \omega^4)}{(4\Phi^2 - \omega^2)^3} q^2,
\end{eqnarray}
\begin{eqnarray}
K_{11}^-({\bf q},\omega;\Phi<E_F) \simeq \frac{E_F}{2\pi} \frac{\omega^2}{4\Phi^2 - \omega^2} - 
\frac{32\Phi^6 - 12 \Phi^2 (3 E_F^2 + \Phi^2)\omega^2 - 3(E_F^2 + \Phi^2)\omega^4}
{12 m^* \pi (4 \Phi^2 - \omega^2)^3} q^2.
\end{eqnarray}
And for $\Phi > E_F$
\begin{eqnarray}
K_{00}^-({\bf q},\omega;\Phi>E_F) \simeq \frac{2m^*}{\pi} \frac{E_F \Phi}{4\Phi^2 - \omega^2} + 
\frac{E_F}{2\pi}\frac{(16 \Phi^3 E_F - 16 \Phi^4 + 12 \Phi E_F \omega^2 + \omega^4)}
{(4\Phi^2 - \omega^2)^3} q^2,
\end{eqnarray}
\begin{eqnarray}
K_{11}^-({\bf q},\omega;\Phi>E_F) \simeq 
\frac{E_F}{2\pi} \frac{4(\Phi E_F - \Phi^2) + \omega^2}{4\Phi^2 - \omega^2}
+ \frac{E_F^2(32 \Phi^3 E_F -48 \Phi^4 + 24 \Phi E_F \omega^2 + 3 \omega^4)}{6 m^* \pi (4 \Phi^2 - \omega^2)^3} q^2.
\end{eqnarray}

\twocolumngrid

\bibliography{references}

\begin{thebibliography}{36}
\bibitem [{spi()}]{spin_note}%
  \BibitemOpen
  \href@noop {} {}\bibinfo {note} {While partial spin-polarization plays an
  important role in the experimentally observed compressible-incompressible
  transition in $\nu_{tot} = 1$ bilayers (see, e.g.,
  Refs.~\onlinecite{spielman05,kumada05,giudici08,zou10}), this transition has
  also been observed in a fully spin-polarized system
  (Ref.~\onlinecite{finck10})}\BibitemShut {NoStop}%
\bibitem [{koh()}]{kohn_note}%
  \BibitemOpen
  \href@noop {} {}\bibinfo {note} {From Kohn's theorem\cite{kohn61} we know the
  effective mass that determines the energy of the in-phase cyclotron mode is
  the bare band mass, not the renormalized mass set by the intralayer Coulomb
  energy. This is a well-known issue in the theory of composite particles which
  is not captured by the RPA.\cite{halperin93,murthy03} As stated in
  Sec.~\ref{secII}, in the present work we take the model under consideration
  and its RPA analysis at face value, even though it is plagued by the usual
  effective mass problems of composite-particle theories.}\BibitemShut {Stop}%
\end{thebibliography}

\end{document}